\documentclass[twocolumn, tighten, times]{aastex631}
\usepackage{graphicx}
\usepackage{hyperref}
\usepackage{epstopdf}
\usepackage{bm}

\usepackage{color}
\usepackage{soul}
\usepackage{threeparttablex, tablefootnote}
\shorttitle{Classification of COSMOS}
\shortauthors{Zhu. ET AL}
\graphicspath{{./}{figures/}}

\begin{document}

\title{Dual-coding contrastive learning based on ConvNeXt and ViT models for morphological classification of galaxies in COSMOS-Web}

\author[0009-0004-0966-6439]{Shiwei Zhu}
\affil{School of Mathematics and Physics, Anqing Normal University, Anqing 246011, China; 
\url{wen@mail.ustc.edu.cn}} 
\affil{Institute of Astronomy and Astrophysics, Anqing Normal University, Anqing 246133, China}

\author[0000-0001-9694-2171]{Guanwen Fang}
\altaffiliation{Corresponding author: Guanwen Fang}
\affil{School of Mathematics and Physics, Anqing Normal University, Anqing 246011, China; 
\url{wen@mail.ustc.edu.cn}} 
\affil{Institute of Astronomy and Astrophysics, Anqing Normal University, Anqing 246133, China}

\author[0000-0002-5133-2668]{Chichun Zhou}
\affil{School of Engineering, Dali University, Dali 671003, China}

\author[0000-0002-0846-7591]{Jie Song}
\affil{Department of Astronomy, University of Science and Technology of China, Hefei 230026, China; \url{xkong@ustc.edu.cn}} 
\affil{School of Astronomy and Space Science, University of Science and Technology of China, Hefei 230026, China}
\affil{Institute of Deep Space Sciences, Deep Space Exploration Laboratory, Hefei 230026, China}

\author[0000-0001-8078-3428]{Zesen Lin}
\affil{Department of Physics, The Chinese University of Hong Kong, Shatin, N.T., Hong Kong S.A.R., China}

\author[0000-0002-4638-0235]{Yao Dai}
\affil{Shanghai Astronomical Observatory, Chinese Academy of Sciences, 80 Nandan Road, Shanghai 200030, China}
\affil{School of Astronomy and Space Science, University of Chinese Academy of Sciences, No. 19A Yuquan Road, Beijing 100049, China}

\author[0000-0002-7660-2273]{Xu Kong}
\affil{Department of Astronomy, University of Science and Technology of China, Hefei 230026, China; \url{xkong@ustc.edu.cn}} 
\affil{School of Astronomy and Space Science, University of Science and Technology of China, Hefei 230026, China}
\affil{Institute of Deep Space Sciences, Deep Space Exploration Laboratory, Hefei 230026, China}

\begin{abstract}
In our previous works, we proposed a machine learning framework named \texttt{USmorph} for efficiently classifying galaxy morphology. In this study, we propose a self-supervised method called contrastive learning to upgrade the unsupervised machine learning (UML) part of the \texttt{USmorph} framework, aiming to improve the efficiency of feature extraction in this step. The upgraded UML method primarily consists of the following three aspects. (1) We employ a Convolutional Autoencoder to denoise galaxy images and the Adaptive Polar Coordinate Transformation to enhance the model's rotational invariance. (2) A pre-trained dual-encoder convolutional neural network based on ConvNeXt and ViT is used to encode the image data, while contrastive learning is then applied to reduce the dimension of the features. (3) We adopt a Bagging-based clustering model to cluster galaxies with similar features into distinct groups. By carefully dividing the redshift bins, we apply this model to the rest-frame optical images of galaxies in the COSMOS-Web field within the redshift range of $0.5 < z < 6.0$. Compared to the previous algorithm, the improved UML method successfully classifies 73\% galaxies. Using the GoogleNet algorithm, we classify the morphology of the remaining 27\% galaxies. To validate the reliability of our updated algorithm, we compared our classification results with other galaxy morphological parameters and found a good consistency with galaxy evolution.  
Benefiting from its higher efficiency, this updated algorithm is well-suited for application in future China Space Station Telescope missions.
\end{abstract}

\section{Introduction} \label{sec:1}

Studying the morphology of galaxies helps in understanding their formation and evolution (\citealt{1958MeLuS.136....1H,1980ApJ...236..351D,10.1093/mnras/sts682}). Galaxies exhibit great diversity in morphology and, particularly, the proportion of irregular galaxies at high redshifts is gradually increasing (\citealt{1996ApJS..107....1A,10.1093/mnras/sts682, Ferreira_2020}).  
Given that galaxy morphology is closely related to numerous physical properties such as color, gas content, star formation rate, stellar mass, and environment (e.g., \citealt{kauffmannEnvironmentalDependenceRelations2004, omandConnectionGalaxyStructure2014, schawinskiGreenValleyRed2014, kawinwanichakijEffectLocalEnvironment2017,guMorphologicalEvolutionAGN2018, lianouDustPropertiesStar2019}), it is essential to have accurate morphological measurements for large galaxy samples.
In previous studies, the most commonly used method is expert visualization (\citealt{1926ApJ....64..321H, 1959HDP....53.....F, Bergh1976ANC}), for example, the Galaxy Zoo as a citizen science project involving multi-person collaborations 
 (\citealt{2011MNRAS.410..166L, 2021AAS...23811902W}). While these programs are popular, they can be inefficient, slow, and subjective, making them poorly suited for future large-scale sky surveys.

With the development of many of the large sky surveys now available, astronomical data has increased dramatically. Several large-scale galaxy mapping projects have been initiated or completed, such as the Sloan Digital Sky Survey (SDSS; \citealt{2002AJ....123..485S}) and the Cosmic Evolution Survey (COSMOS; \citealt{Scoville_2007}). These projects have significantly increased the number of observed galaxies (\citealt{2006ApJ...652..963R, Scoville_2007}). Since traditional methods are unable to efficiently handle such large astronomical datasets, the application of machine learning, especially deep learning techniques, in this field is particularly important.

In recent years, convolutional neural networks (CNNs; \citealt{SCHMIDHUBER201585}) have been widely used to solve galaxy morphology-related problems,
such as galaxy-star classification (e.g., \citealt{Bai_2018, Zhang_2024}), identification of galaxy mergers (e.g., \citealt{10.1093/mnras/stab1677, 2024MNRAS.531.4070D}), classification of galaxy morphology, and searching for gravitational lensing (e.g., \citealt{10.1093/mnras/stac562, 2024MNRAS.tmp.1656M}). Since CNN is a supervised machine learning (SML) method, it requires a large number of pre-labeled datasets as training sets, which are usually obtained by visual inspection. In the face of the large amount of astronomical data obtained from large-scale sky survey programs, this approach is time inefficient  and does not take full advantage of 
machine learning.

The unsupervised machine learning (UML) method provides a solution to classify the galaxy morphology based on different features without pre-labeling. This method is usually divided into two steps: (1) extracting features from the original images, and (2) clustering galaxies based on similar features. UML methods have been widely used due to their powerful data processing and transfer learning capabilities
(e.g., \citealt{10.1093/mnras/stx2351, 10.1093/mnras/stab734, Tohill_2024}). For example, \cite{Tohill_2024} 
used Variational Autoencoders (VAEs) to perform feature extraction and classification of galaxy morphologies at $2 < z < 8$ observed by the James Webb Space Telescope (JWST; \citealt{SCHMIDHUBER201585}).

Self-supervised learning (SSL) is also effective in handling unlabeled data, where contrastive learning (e.g., \citealt{Chen2020ASF,Chen2020ImprovedBW}) stands out as a predominant direction in SSL. The essence of this method lies in constructing positive and
negative sample pairs, allowing the model to learn to distinguish between similar samples without the necessity of external labels. 
Traditional contrastive learning primarily constructs positive pairs through data augmentation techniques. For example,  For example, SimCLR \citep{Chen2020ASF} generates positive samples by applying geometric transformations to an image, with different images serving as negative samples. In contrast, BYOL \citep{Grill2020BootstrapYO} does not rely on negative samples and forms positive pairs by predicting the representation of the same image in different augmented views. MOCO \citep{Chen2020ImprovedBW} constructs dynamic dictionaries to ensure that query vectors are similar to key vectors. SwAV \citep{caron2021unsupervised} combines UML clustering to reinforce the consistency between cluster assignments for different views of the same image. These algorithms require a certain level of task-specific knowledge to determine which key features should be retained as positive samples. After constructing positive and negative samples, contrastive learning demonstrates stronger feature extraction capabilities, thereby enhancing the model's classification performance.

In previous works, \cite{Zhou_2022} innovatively developed a Bagging-based Voting Clustering Method to obtain 100 samples with high inter-group similarity while discarding some of the data. Following this, 
\cite{Fang_2023} introduced the Adaptive Polar Coordinate Transformation (APCT) technique, which improves the robustness and performance of the model by enhancing the rotational invariance of images, and their experiments demonstrated that GoogLeNet performs exceptionally well in galaxy classification. \cite{Dai_2023} combined the UML and SML models which successfully classified the full dataset of 17,292 galaxies from the COSMOS-DASH field and provided a reliable catalog.
\cite{Song_2024} proposed the \texttt{USmorph} model by combining the UML and SML methods, expanding the dataset sixfold and demonstrating the efficiency and reliability of the combined UML and SML model in classification.
\cite{Fang2024} enhanced the UML step by applying Convolutional Neural Network (ConvNeXt; \citealt{10205236}) large model encoding and Principal Components Analysis (PCA; \citealt{MACKIEWICZ1993303}) dimensionality reduction techniques, achieving a classification effect of 20 categories.
Our series of studies have demonstrated that the classification framework combining UML and SML possesses sufficient robustness and efficiency to adapt to different sky regions. It overcomes the reliance on single supervised algorithms on data labels and their limited transfer learning capabilities.

Based upon our prior research, the USmorph architecture has been enhanced with the following advancements: (1) after data preprocessing, feature extraction is implemented using a hybrid ConvNeXt and Vision Transformer (ViT; \citealt{Dosovitskiy2020AnII}) model structure; (2) a 
secondary feature extraction process is then applied to achieve effective dimensionality reduction by leveraging contrastive learning. We adopt the pre-trained ConvNeXt and ViT models to encode images for positive and negative sample pair construction. This approach does not rely on specific data augmentation techniques, making it applicable to a wider range of tasks and datasets and thus enhancing its generalization capability. During this process, the encoders can extract key features from the data. Subsequently, by comparing all positive and negative samples, the critical features of the data are highlighted, while redundant information is eliminated, achieving the effect of dimensionality reduction. After that, we can use the Bagging-based clustering model \citep{Zhou_2022} to classify the encoded data into 10 groups, which greatly improves the classification efficiency from 100 groups \citep{Zhou_2022} to 10 groups.
We conducted SML training and visualization tests using Uniform Manifold Approximation and Projection (UMAP; \citealt{McInnes2018UMAPUM}) based on the labels obtained through the Contrastive learning method. Additionally, we evaluated our labels against morphological parameters, demonstrating the reliability of our classification results.

The structure of this paper is as follows. Section \ref{sec:2} provides an overview of the COSMOS-Web program and sample selection criteria. Section \ref{sec:3} introduces the Contrastive learning method. Section \ref{sec:4} presents classification results and system parameter measurements. Section \ref{sec:5} summarizes the main conclusions and provides an outlook for future research. Throughout this paper, we use the AB magnitude system \citep{1983ApJ...266..713O} and assume a \cite{Chabrier_2003} initial mass function and a standard flat $\Lambda$CDM cosmology with parameters $H_0 = 70$ km s$^{-1}$ Mpc$^{-1}$, $\Omega_m = 0.3$, and $\Omega_\Lambda = 0.7$.

\section{Data selection} \label{sec:2}

\subsection{COSMOS-Web}
COSMOS-Web (\citealt{2023ApJ...954...31C}) is a large-scale deep-space observational program conducted by the JWST in its first observational cycle. This project aims to investigate the origin of the Universe and the formation of early galaxies through extensive infrared imaging of a designated region known as the COSMOS field. The COSMOS-Web survey encompasses a contiguous 0.54 deg$^2$ area of the sky, enabling it to image a significant number of high-redshift galaxies. For this purpose, the project utilized the NIRCam \citep{2022SPIE12180E..3PW} and MIRI \citep{Rieke_2023} instruments aboard the JWST. The NIRCam imaging was performed using four distinct infrared filters (F115W, F150W, F277W and F444W), while the MIRI imaging employed a single filter (F770W).

\subsection{COSMOS2020 Catalog}\label{sec:2.2}
The ``Classic'' COSMOS2020 catalog \citep{Weaver_2022} furnishes comprehensive photometric data covering 35 bands, ranging from ultraviolet to near-infrared wavelengths. This catalog utilizes flux data to estimate redshifts, stellar masses, and other physical properties of galaxies through spectral energy distribution (SED) fitting techniques.
For the determination of photometric redshifts, \citet{Weaver_2022} used two different codes, {\tt EAZY} \citep{Brammer_2008} and {\tt LePhare} \citep{Ilbert_2009}, to assess the consistency between their results. Here, we adopted the redshifts obtained from {\tt LePhare}, as shown in Figure~15 of \citet{Weaver_2022}.

The redshift estimation process with {\tt LePhare} involved a library of 33 galaxy templates based on the models of \citet{BruzualCharlot} and \citet{Ilbert_2009}. The analysis also considered various dust extinction and attenuation curves, including the starburst attenuation curve from \citet{Calzetti_2000}, the Small Magellanic Cloud extinction curve from \citet{1984A&A...132..389P}, and two modifications of the \citet{Calzetti_2000} law that account for the 2175\,\AA\ absorption feature. The final photometric redshift ($z_{\text{LePh}}$) was determined as the median value of the redshift likelihood distribution.

\subsection{Sample Selection}
We utilize the NIRCam second epoch data from the COSMOS-Web field, as provided by \cite{2023ApJ...954...31C}, The chosen pixel scale is $0\farcs03$.
Based on the principle of stellar mass completeness \citep{2010A&A...523A..13P}, we selected rest-frame optical band images in four different infrared filters (F115W, F150W, F277W, and F444W) from the COSMOS-Web survey.

We selected the parent sample of galaxies from the COSMOS2020 catalog based on the following criteria:(1) $\rm lp_{type}=0$, to ensure that we select galaxies rather than stars;
(2) $\rm FLAG_{ COMBINE}=0$, which means that flux measurements are unaffected by the bright stars and that the objects are not at the image edges, thus guaranteeing reliable photometric redshift and stellar mass estimations;
(3) $0.5 < z < 6.0$, it is essential to maintain the overall wavelength coverage within the range of 4000 to 8000 \AA, with the central wavelength of the samples in each filter being as close as possible to $\sim 5500$ \AA. 

\begin{figure}
    \includegraphics[width=0.45\textwidth]{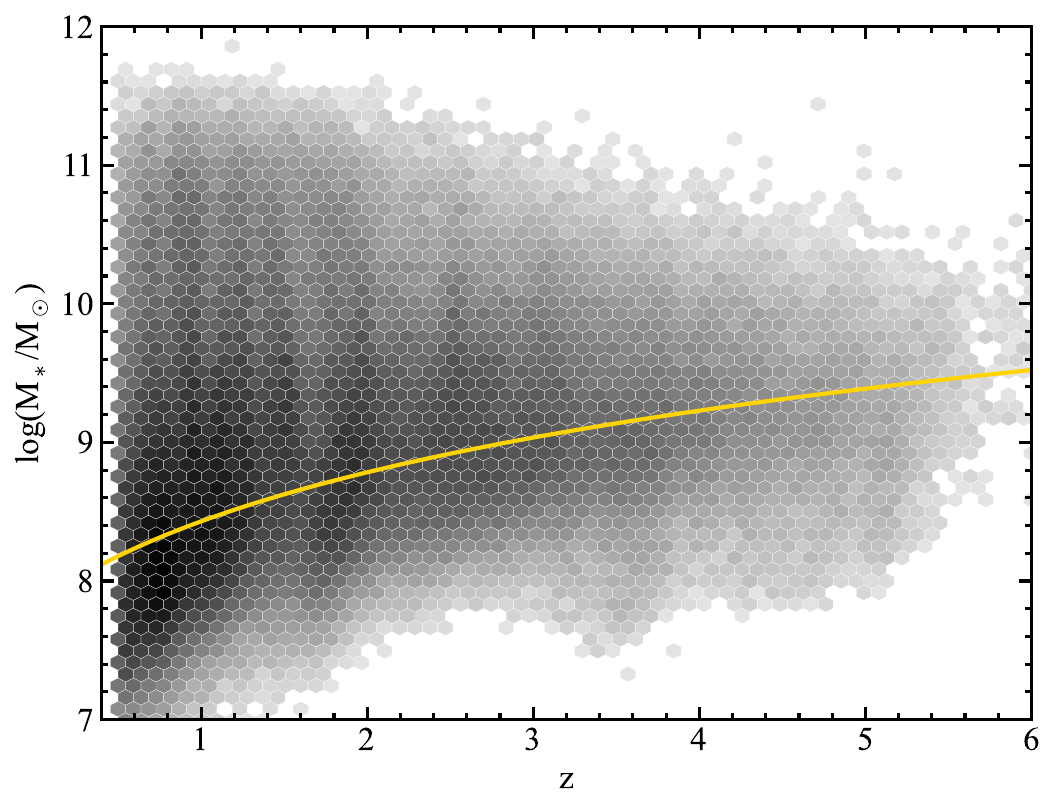}
   \caption{Stellar mass as a function of redshift for the parent sample of this study. 
The stellar-mass completeness, given by $ M_{\rm comp}(z)/M_{\odot} = -1.51 \times 10^6 (1 + z) + 6.81 \times 10^7 (1 + z)^2$, is indicated with the solid yellow line. Background gray dots represent the parent sample selected from the COSMOS2020 catalog (see Section \ref{sec:2.2}).}    
\label{fig:1}
\end{figure}

Since more massive galaxies can be detectable than low-mass galaxies as redshift increases due to the limitations of the observational depth and completeness, the sample variance could be strongly affected {\citep{Bates_2019}}.
To relieve this bias, we took the mass completeness of the parent sample into account by following the method described in \cite{Weaver_2022}. Specifically, \cite{2010A&A...523A..13P} defined a minimum stellar mass ($M_{\rm min}$) at each redshift, above which the selected sample is generally completed. To determine $ M_{\rm min} $, the limiting stellar mass ($ M_{\rm lim} $) of each galaxy needs to be estimated using the completeness limit function given by $ M_{\rm comp}(z)/M_{\odot} = -1.51 \times 10^6 (1 + z) + 6.81 \times 10^7 (1 + z)^2$. In Figure \ref{fig:1}, we illustrate the distribution of stellar mass ($ M_* $) as a function of redshift for galaxies at $0.5 < z < 6.0$ in the COSMOS2020 catalog, which serves as the parent sample of this study. The completeness limit function, $ M_{\text{comp}}(z) $, is depicted by a solid yellow line. Only galaxies exceeding the mass completeness limit are selected for our sample.

Ultimately, we selected a total of 45,288 galaxies in the COSMOS field. The redshift intervals corresponding to the filters are summarized in Table \ref{tab1}. 

\begin{table}[htb!]
\centering
\caption{Summary of the bands from JWST/NIRCam with the associated redshift that aligns with the optical rest-frame.\label{tab1}}
\begin{tabular}{cccc}
\hline\hline
Redshift & Band & $\overline{\lambda}_{\text{rest}}$ (Å) & Total \\
\hline
$0.5 < z \leq 1.2$ & F115W & 6,113 & 18,284 \\
$1.2 < z \leq2.9$ & F150W & 5,191 & 21,535 \\
$2.9 < z \leq5.3$ & F277W & 6,068 & 5,294 \\
$5.3 < z < 6.0$ & F444W & 6,713 & 175 \\
\hline
\end{tabular}
\end{table}

\section{METHOD FOR MORPHOLOGICAL CLASSIFICATION} \label{sec:3}
In this section, we will introduce an updated framework that applies dual-encoding contrastive learning techniques to enhance our UML model.

The framework consists of four main components: (1) data preprocessing, (2) extraction of important features from the data, (3) highlighting important features through comparison, and (4) voting based on a Bagging clustering model. Finally, we utilize the labels obtained from the UML method to train the GoogLeNet model.

\begin{table}[htb!]
\centering
\caption{CAE architecture \label{tab2}}
\begin{tabular}{ccccc}
\hline\hline
Layer & Operation & Dimension & Filter Size  & Stride \\
\hline
L0 & Input       & $100\times 100 \times 1$ & ... & ... \\
L1 & Convolution & $100\times 100 \times 8$ & $5\times 5$ & ... \\
L2 & Maxpooling  & $50 \times 50  \times 8$ & $2\times 2$ & $2\times 2$ \\
L3 & Convolution & $50 \times 50  \times 8$ & $5\times 5$ & ... \\
L4 & Maxpooling  & $25 \times 25  \times 8$ & $2\times 2$ & $2\times 2$ \\
L5 & Unfolding   & 10000   & ... & ... \\
L6 & Full connection & 40 & ... & ... \\
\hline
\end{tabular}
\begin{tablenotes}
\item[]{Note:The CAE architecture: an input layer (L0); two convolutional layers (L1, L3) with 8 filters each, each followed by a 2×2 max-pooling layer (L2, L4) with stride 2; an unfolding layer (L5) flattening the output to 10,000 dimensions; and a fully connected layer (L6) reducing dimensions to 40 for encoded representation.}
\end{tablenotes}
\end{table}

\begin{figure*}
    \includegraphics[width=2\columnwidth]{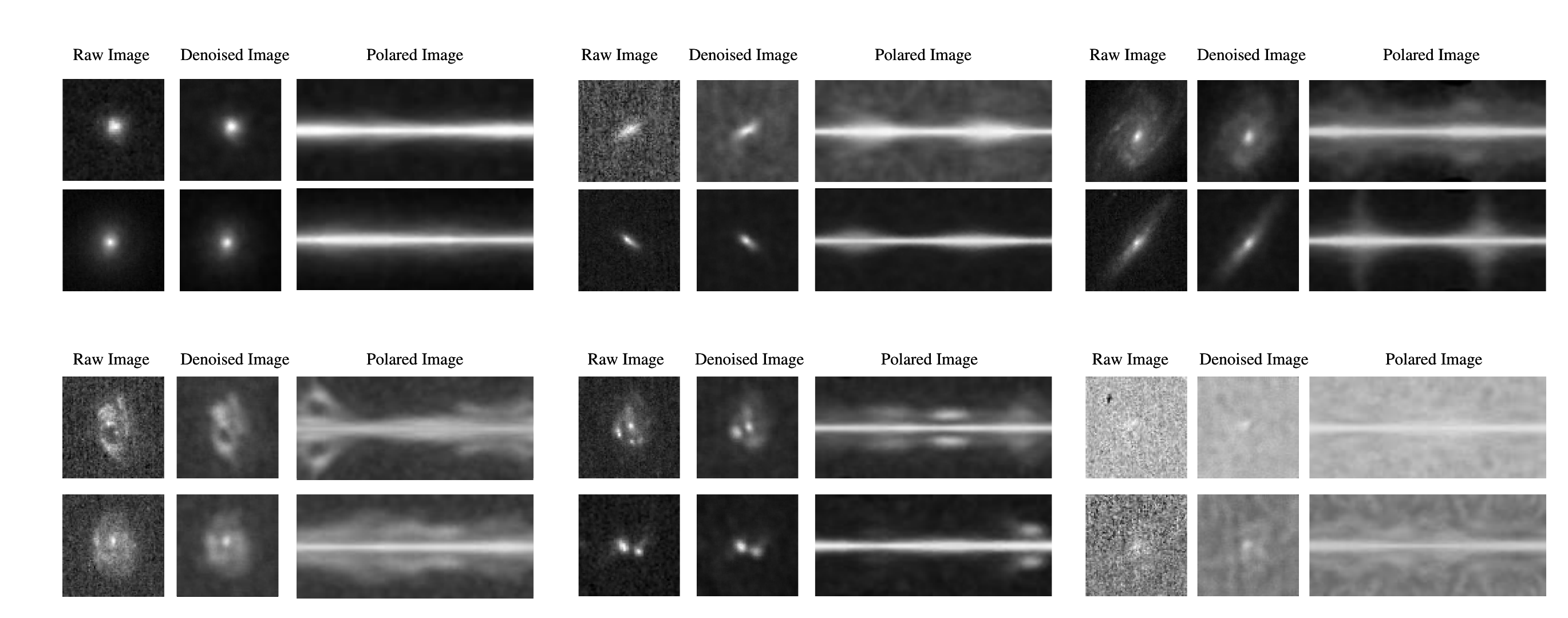}
    \caption{Comparisons between the original and preprocessed galaxy images for six different morphological types. In each case, images are arranged in a sequence from left to right: original image, denoised image, and image after the polar coordinate transformation.}
    \label{fig:2}
\end{figure*}

\begin{figure}
\centering 
  \includegraphics[width=\columnwidth]{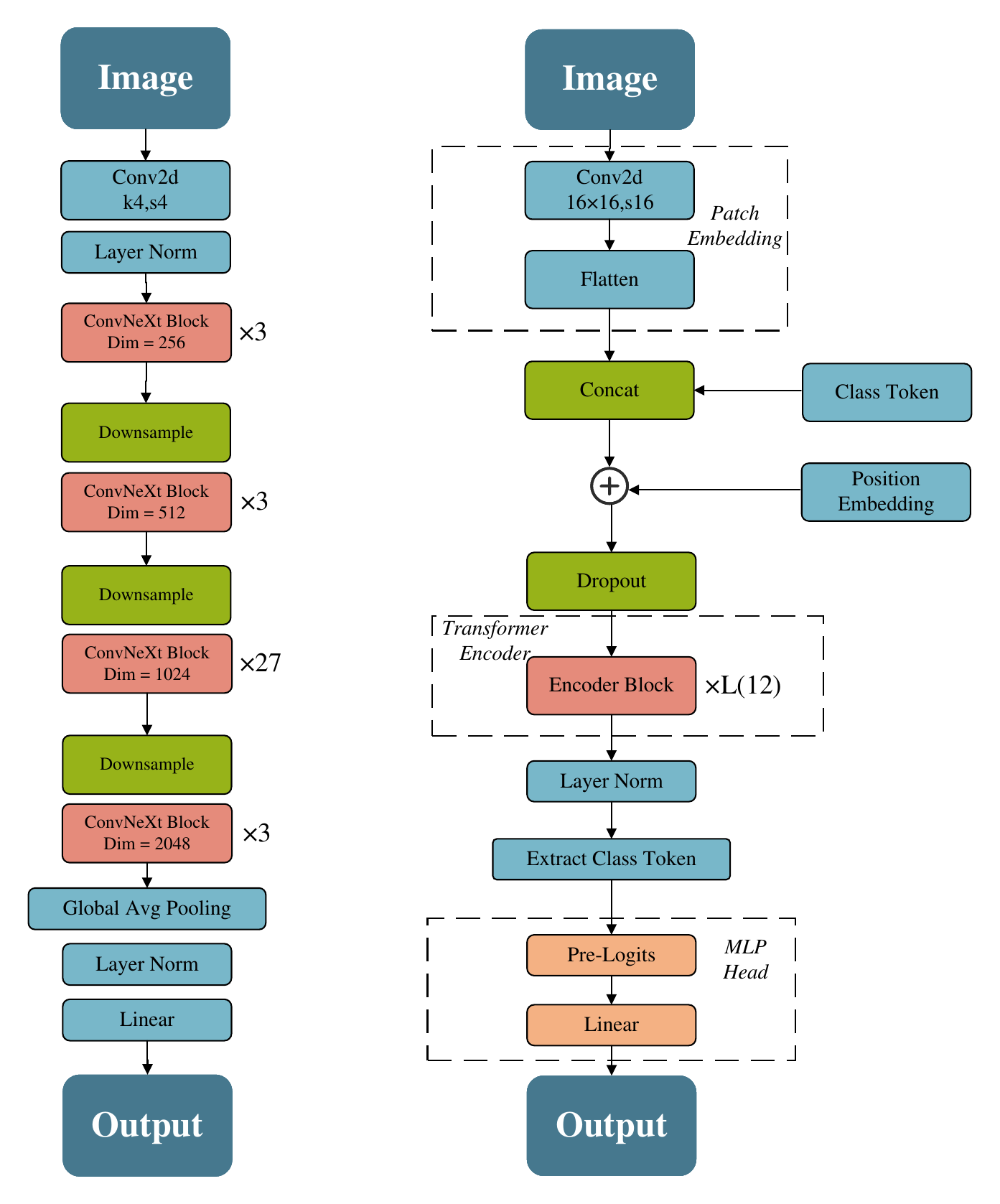}
    \caption{Frameworks of the ConvNeXt (left) and ViT (right) models.}
\label{fig:3}
\end{figure}    

\subsection{Data pre-processing}
We adopted the approach for cropping image sizes based on the work of \cite{Song_2024}. Given that our study encompasses a wider redshift range than \cite{Song_2024}, we adjusted our image cropping size to $75 \times 75$ pixels. By utilizing {\tt\string GALAPAGOS}  software (\citealt{Barden2012GALAPAGOSFP,2022A&A...664A..92H}), we measured the effective radii of galaxies within our sample and discovered that 95\% of them have an effective radius smaller than 40 pixels. This adjustment ensures that our cropped images retain adequate galactic information for effective morphological classification. 

To enhance the classification performance of the model, a two-step image processing method was employed to address the issues of image noise reduction and the mitigation of adverse effects caused by poor rotational robustness. As illustrated in the initial steps of Figure \ref{fig:2}, the detailed procedure is as follows.
To tackle the negative impacts on classification results caused by a low signal-to-noise ratio (S/N) and high levels of noise in images \citep{liu2022networkspixelsembeddingmethod}, we implemented a noise reduction technique on the image data while ensuring the preservation of primary features. The Convolutional Autoencoder (CAE; \citealt{10.1007/978-3-642-21735-7_7}) excels in noise reduction and feature extraction. It automatically identifies key features within images via convolution and pooling operations and utilizes these features to reconstruct the denoised images \citep{Zhou_2022, Fang_2023, Dai_2023, Song_2024,Fang2024}. As depicted in Figure \ref{fig:3}, the left image of each example galaxy represents the original image, whereas the center image illustrates the reconstructed image post-CAE denoising. This comparison highlights that the CAE can effectively reduce data noise while preserving the core information of the images via extracting significant features from the galaxy pictures. Table \ref{tab2} outlines the convolution kernel sizes and architecture of the CAE utilized in this study. Experimental determination indicated that the model performs optimally with a convolution kernel size of 5$\times$5.

Image rotation processing might cause misclassification in the galaxy morphology classification task (\citealt{8899285}). To ensure that the classification of galaxy morphology is rotationally invariant, we employ a technique called APCT \citep{Fang_2023}.

This approach not only effectively enhances the model's rotational invariance but also markedly reduces computing costs compared to conventional data augmentation techniques such as rotation, masking, and sorting. The technique is based on defining the polar axis by identifying the brightest and darkest pixels in the image, rotating the polar axis counterclockwise in steps of 0.05 rad each time, and then ``stacking'' all the pixels along the rotated polar axis into polar coordinates. Finally, the image is mirrored to emphasize the central features, which enhances the morphological features of the image and the rotational invariance of the model. In the right column of Figure \ref{fig:3}, we show the corresponding results after the implementation of APCT.

\begin{figure*}
    \includegraphics[width=2\columnwidth]{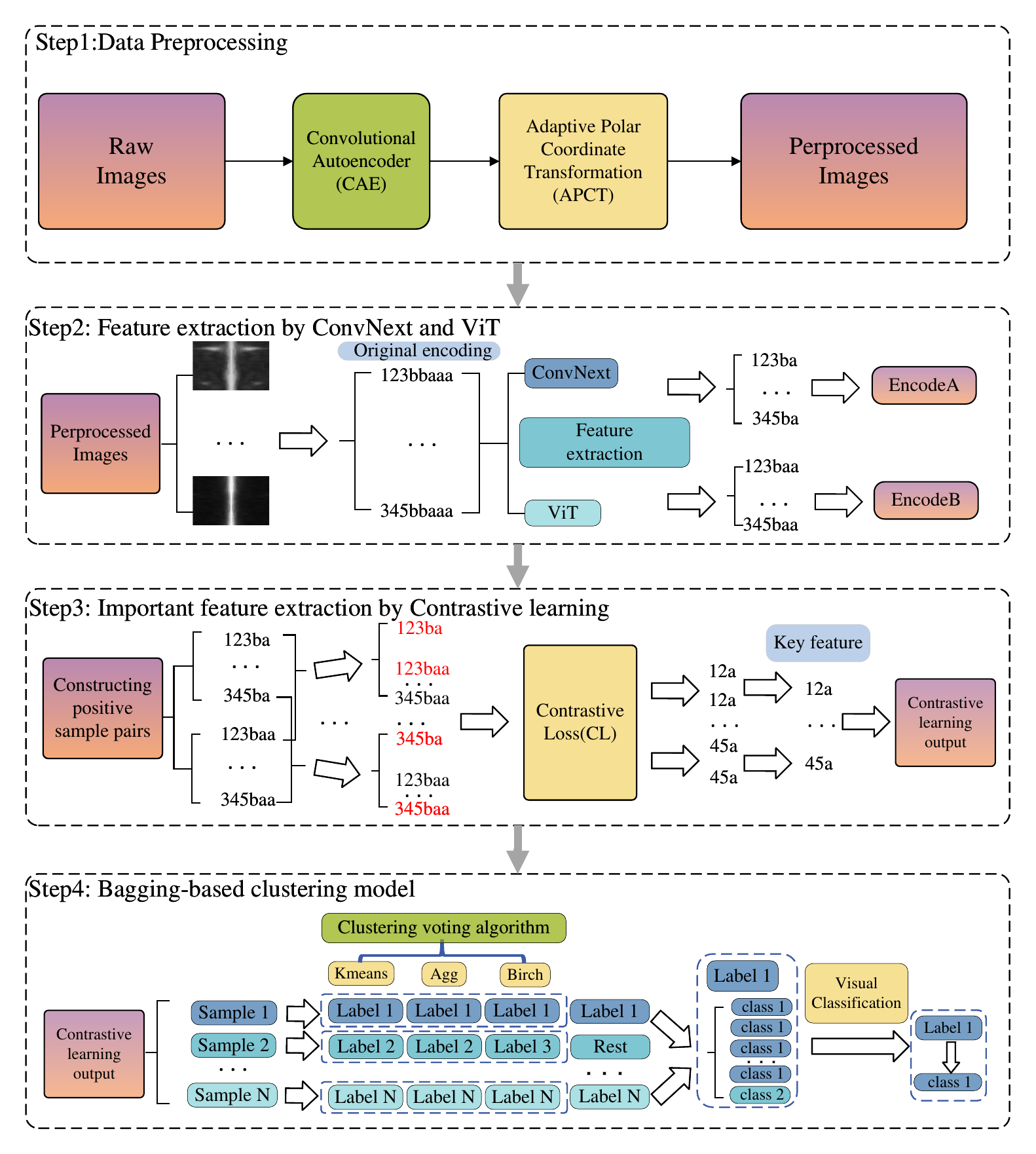}
    \caption{Schematic diagram of UML Clustering Process:  Upon completing data preprocessing (i.e., Step1), we employ both the ConvNeXt model and the ViT model to encode the data, thereby extracting compelling features  (see Step2). The encoding results from identical data instances are used to construct positive sample pairs, while the remaining encodings from the ViT model serve to form a negative sample set. We subsequently apply the Contrastive Loss (CL) method to minimize loss, which helps in identifying critical features (Step3). For the final determination of labels in Step4, we adopt a voting mechanism based on a Bagging clustering model, supplemented by manual visual adjustments to ensure precision. This approach leverages the strengths of two advanced models, integrating automated learning with human oversight to deliver a precise and reliable feature identification and classification solution.
}
    \label{fig:4}
    
\end{figure*}

\subsection{Feature Extraction and Contrastive Learning} \label{subsec:morph}
Pre-trained visual classification models are first trained on large general-purpose image datasets to obtain good initial parameters and feature representations, which can then be fine-tuned for use in specific tasks. Such models can efficiently extract features even without input data labels. Pre-trained models represented by ConvNeXt (\citealt{10205236}) and ViT (\citealt{Dosovitskiy2020AnII}) have been widely used as feature encoders in image classification with remarkable results \citep{liu2023simple, Fang2024}.
The primary mechanism of contrastive learning involves constructing positive sample pairs and negative sample set. Within this framework, positive samples (similar samples) are drawn closer together in the feature space, whereas negative samples (dissimilar samples) are pushed further apart. This spatial adjustment aids in extracting fundamental features from the data. A significant challenge in traditional contrastive learning has been the efficient construction of high-quality positive sample pairs (e.g., \citealt{Chen2020ASF,Chen2020ImprovedBW,Grill2020BootstrapYO,caron2021unsupervised}).

To address the challenge in traditional contrastive learning methods that rely on prior knowledge for selecting data augmentation techniques when constructing positive and negative sample pairs, we utilize the encoding outputs from pre-trained large-scale supervised models, specifically ConvNeXt and ViT, to build our positive-negative sample modeling framework (see Figure \ref{fig:3}). This approach not only reduces the reliance on prior knowledge but also enhances the capability of feature extraction, thereby making our method applicable to a wide range of datasets. Each piece of sample data is encoded using both ConvNeXt and ViT models.

Subsequently, due to the superior image extraction capabilities demonstrated by the ConvNeXt model in previous works, it has been selected as the anchor sample (\citealt{liu2023simple,huang2025unsupervisedwasteclassificationdualencoder}), which outputs a feature dimension of 2048, in contrast to the ViT model's output of 1000 dimensions, we employed a linear layer to project the 1000-dimensional output of the ViT model to 2048 dimensions. This adjustment ensures dimensional consistency between the two models within the dual-encoder contrastive learning framework, facilitating a fair comparison under identical dimensionality. Such alignment is crucial for effective similarity measurement and contrastive loss computation, thereby enabling the learning of meaningful feature representations.

\begin{equation}
    EncodeA = ConvNeXt(x_1, x_2, x_3, ... , x_n)
\end{equation}
\begin{equation}
    EncodeB = Linear(ViT(x_1, x_2, x_3, ... , x_n))
\end{equation}
where \(x_{(x_1, x_2, x_3, ... , x_n)}\) denotes the set of pixel matrices of the input images, \(ConvNeXt()\) and \(ViT()\) denote the encoder type, and \(Linear()\) denotes the linear layer.

Upon completion of the encoding process, the construction of positive and negative sample pairs commences. To ensure effective contrastive learning, lower-quality feature encodings are utilized as negative samples. This approach enhances the differentiation of key features by providing a clearer contrast against higher-quality encodings. 
For the same galaxy $g$, we take $(\text{EncodeA}(g), \text{EncodeB}(g))$ as a positive sample pair. The encoding results of other galaxies $ g' $ ($ g'\neq g $) representing as $ \text{EncodeB}(g') $ form the negative sample set. Each positive sample pair is matched with all samples in the negative set.
By comparing these positive pairs and all negative samples and under the guidance of a contrastive learning loss function (Contrastive Loss, CL), a nonlinear layer is used to establish the mapping relationship between sample encodings and critical features.
The nonlinear layer learns this mapping by minimizing the CL. A smaller CL value signifies that the output of the nonlinear layer is closer to the key features of the target data, effectively ensuring that samples of the same class are more closely clustered in the feature space, while samples from different classes are pushed further apart.

The CL is defined as follows:
\begin{equation}
CL = \frac{1}{2N} \sum_{i=1}^{n} loss_i
\end{equation}
Where \(N\) is the amount of input data, \(loss_i\) is the loss function between each pair of positive and negative samples, and \(i\) is the first \(i\) data out of the input \(N\) data that is selected as a positive sample after Eqs. (1) and (2). \(loss_i\) is defined as follows:
\begin{equation}
loss_i = -\log \frac{\exp(S_{ij}/\tau)}{\exp(S_{ij}/\tau) + \sum_{k \neq i} \exp(S_{ki}/\tau)},
\end{equation}
in which $S_{ij} = sim(z_i, z_j)$ is the cosine similarity between the positive sample \( z_i \) (obtained by processing the selected data with Equation (1)) and the negative samples \( z_j \) (obtained by processing all other data with Equation (2)), and \( \tau \) is the temperature parameter.
The cosine similarity \( sim(z_i, z_j) \) is defined as:

\begin{equation}
sim(z_i, z_j) = \frac{z_i \cdot z_j}{\|z_i\| \cdot \|z_j\|}
\end{equation}
where \( z_i \cdot z_j \) is the dot product between the positive and negative samples, and \( \|z_i\| \) and \( \|z_j\| \) are the magnitudes (norms) of the vectors \( z_i \) and \( z_j \), respectively.

As shown in Step2 and Step3 of Figure \ref{fig:4}, the contrastive learning based on ConvNeXt and Vit outputs can retain the main features and remove the redundant features for our subsequent processing.

\subsection{UML Clustering Process}
After completing data preprocessing and pairwise dimensionality reduction, we adopt a UML-based classification method proposed by \cite{Zhou_2022} for classifying galaxy samples. The specific procedure involves implementing a Bagging-based multi-model voting classification approach.

We utilized three different clustering models for data processing, each of which designed to classify the sample images into ten categories: the Hierarchical Balanced Iterative Reduction and Clustering Algorithm \citep{10.1145/233269.233324, Peng2020}; the K-means Clustering Algorithm \citep{8ddb7f85-9a8c-3829-b04e-0476a67eb0fd}; and the Hierarchical Agglomerative Clustering Algorithm \citep{10.1093/comjnl/26.4.354, Murtagh_2014}. We adopted the labels of the K-means algorithm as the dominant labels and assigned the corresponding labels to the other clustering models based on the frequency of the K-means labels within each group, and implemented a majority voting filtering strategy. When all three algorithms generate identical voting outcomes, their intersection constitutes the final sample set. Any samples falling outside this set are discarded. After discarding 27\% galaxies as the disputed samples, we obtained a highly robust set of samples. The clustering process refers to Step 4 in Figure \ref{fig:4}.

\begin{figure*}	\includegraphics[width=2\columnwidth]{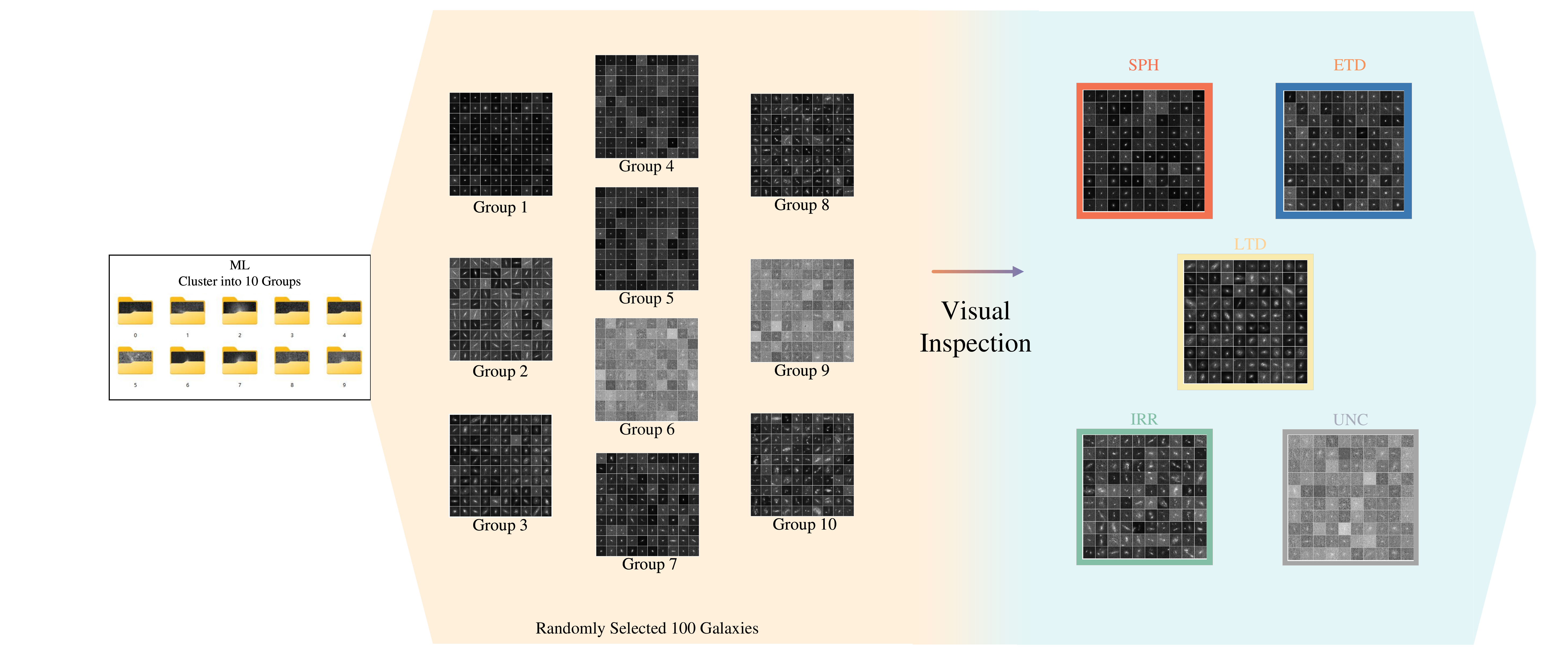}
    \caption{Randomly select 100 images from each of the 10 machine clusters for display. We presented all 10 clustering results, which are characterized by small differences in the morphology of galaxies within the group and large differences in the morphology of galaxies outside the group. Therefore, we can visually classify each component into one of the five types (SPH, ETD, LTD, IRR, and UNC). Due to the limited number of categories, the visual efficiency is greatly improved.}
    \label{fig:5}
\end{figure*}

Following the removal of 12,366 samples, we divided the remaining 32,922 sources into 10 groups. Each of these groups exhibited minimal internal variation and significant differences between groups. Following the methodology outlined by \cite{liu2023simple}, we only need to visually inspect 100 randomly selected galaxy images from each group to obtain the corresponding labels. To minimize subjective bias, three trained experts conducted a voting process in this visual inspection step. A final label was determined only when at least two experts agreed. This process is illustrated in Figure \ref{fig:5}. Ultimately, we classified the 32,922 galaxies into five distinct morphological categories: spherical (SPH), early-type disk (ETD), late-type disk (LTD), irregular (IRR), and unclassified (UNC). The specific distribution of numbers for each category can be found in Table \ref{tab3}.

\subsection{SML Clustering Process}
After removing 12,366 sources with inconsistent voting results, we obtained a set of 32,922 galaxies with reliable morphological labels through the UML clustering process. These galaxies were used as the training set for the SML analysis of the remaining 12,366 galaxies. According to \cite{Fang_2023}, GoogLeNet has shown excellent performance in deep-field galaxy classification. Therefore, we use the GoogLeNet algorithm \cite{7298594} as a supervised classification model.

\begin{figure*}	\includegraphics[width=2\columnwidth]{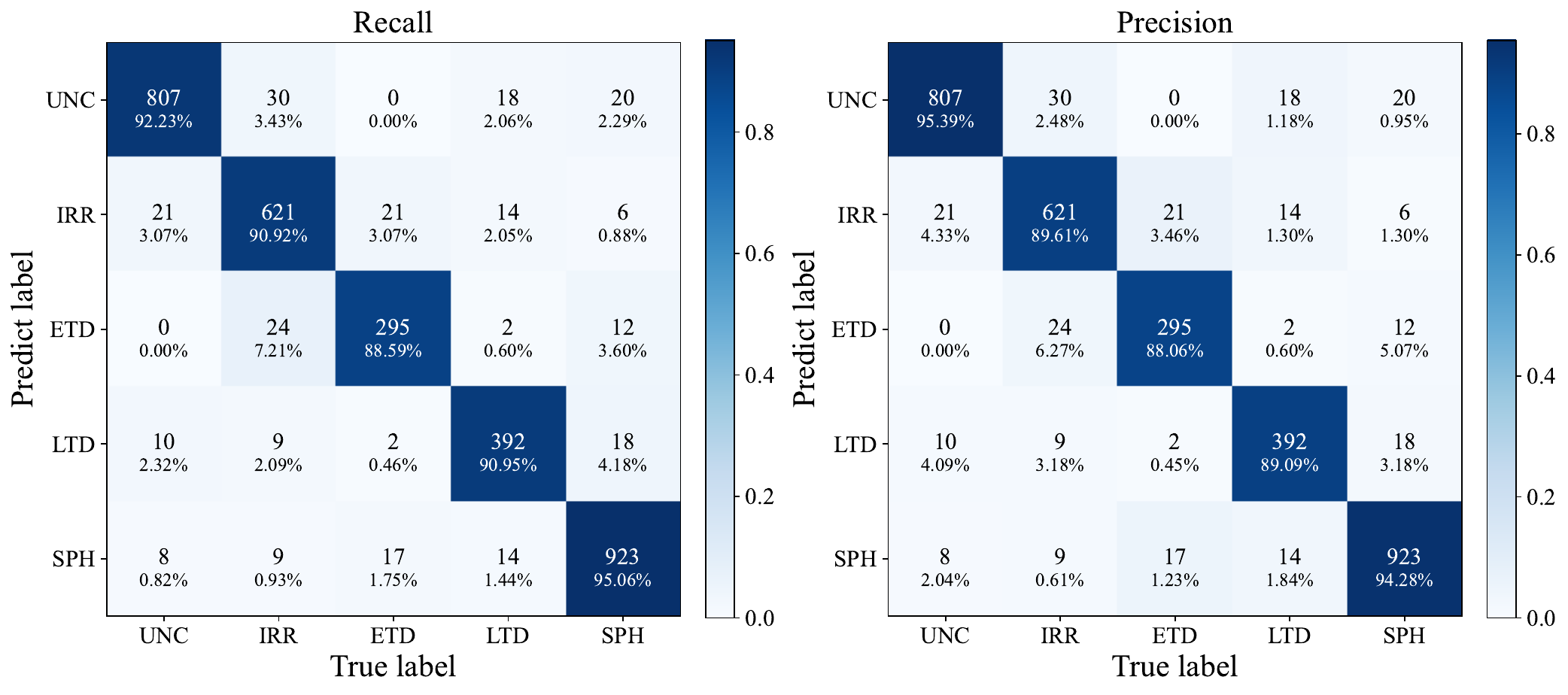}
    \caption{Recall (left) and precision (right) of the GoogLeNet model. Note that both metrics have an overall average exceeding 92\%, indicating the model's excellent performance in distinguishing between different galaxies.}
    \label{fig:6}
\end{figure*}

To avoid overfitting, we adopted the same setup as described in \cite{Fang_2023}, dividing the labeled data obtained from the UML step randomly into a training set (32,922 samples) and a validation set (3,293 samples) at a fixed ratio of approximately 9:1. As illustrated in Figure \ref{fig:6}, our results indicate that the GoogLeNet model exhibits superior performance, achieving an overall precision and recall exceeding 92\%, which underscores its robustness and reliability for galaxy classification. The distribution of classification results using the SML method is also listed in Table \ref{tab3}.

Finally, we conducted a horizontal comparison of the dual-encoding contrastive learning framework with the original {\tt\string USmorph} in terms of performance. Under the same COSMOS-Web dataset, the new framework demonstrated its superiority in overall reject, accuracy, and recall rate, as shown in Table \ref{tab4}.

\textbf{}
\begin{table}[htb!]
\center
\caption{Number of galaxies in different morphological types\label{tab3}}
\begin{tabular}{ccccccc}
\hline\hline
Model & SPH & ETD & LTD  & IRR & UNC& TOTAL\\
\hline
UML & 9,942 &3,255 & 4,076 & 6,952& 8,697 & 32,922\\
SML & 4,253 &731& 3,128 & 1,994& 2,260 & 12,366\\
TOTAL& 14,195 &3,986 & 7,204 & 8,946& 10,957 & 45,288\\
\hline
\end{tabular}
\end{table}

\begin{table}[htb!]
\center
\caption{Comparison of classification performance of different methods\label{tab4}}
\begin{tabular}{ccccccc}
\hline\hline
Method & dataset count & Reject& Accuracy  & Recall \\
\hline
Our Method  & 45,288 &27.31\% & 92.26\% & 92.25\%\\
{\tt\string USmorph} & 45,288 & 39.61\% & 74.86\% & 74.87\% \\

\hline
\end{tabular}
\end{table}

\begin{figure*}
\includegraphics[width=2\columnwidth]{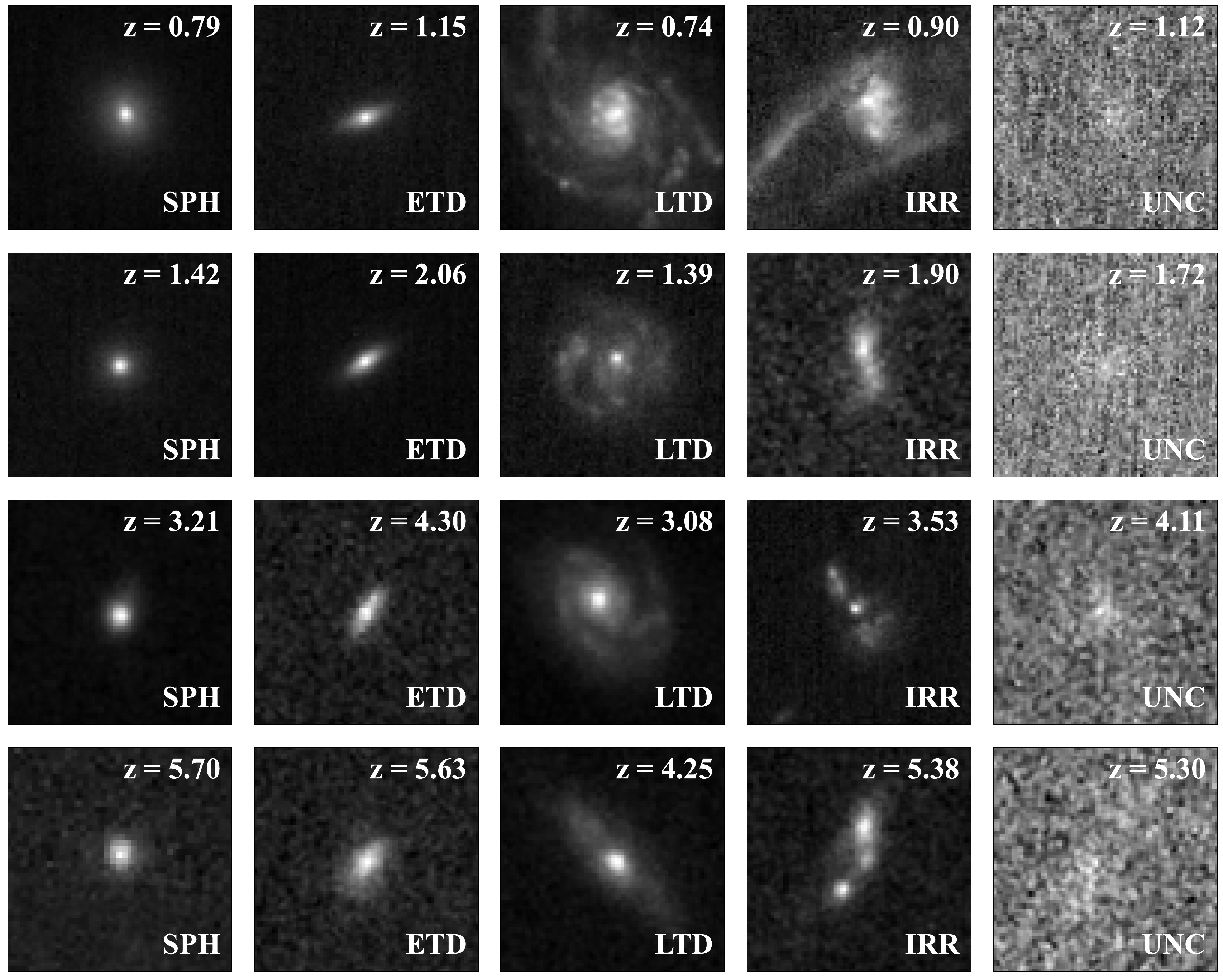}
    \caption{Exampled galaxies belonging to the five morphological classes: SPHs, ETDs, LTDs, IRRs, and UNCs, from left to right. SPH galaxies are characterized by a distinct spherical shape and centrally concentrated brightness. ETD galaxies feature a prominent central sphere and a relatively smaller disk-like structure. LTD galaxies are defined by their well-defined spiral arms and a compact central sphere. IRR galaxies have an irregular structure, while UNC galaxies are affected by an extremely low signal-to-noise ratio, which makes it difficult to identify the embedded information. Redshifts of individual galaxies are also given.}
    \label{fig:7}
\end{figure*}

\begin{figure*}
    \centering
    \includegraphics[width=0.85\textwidth]{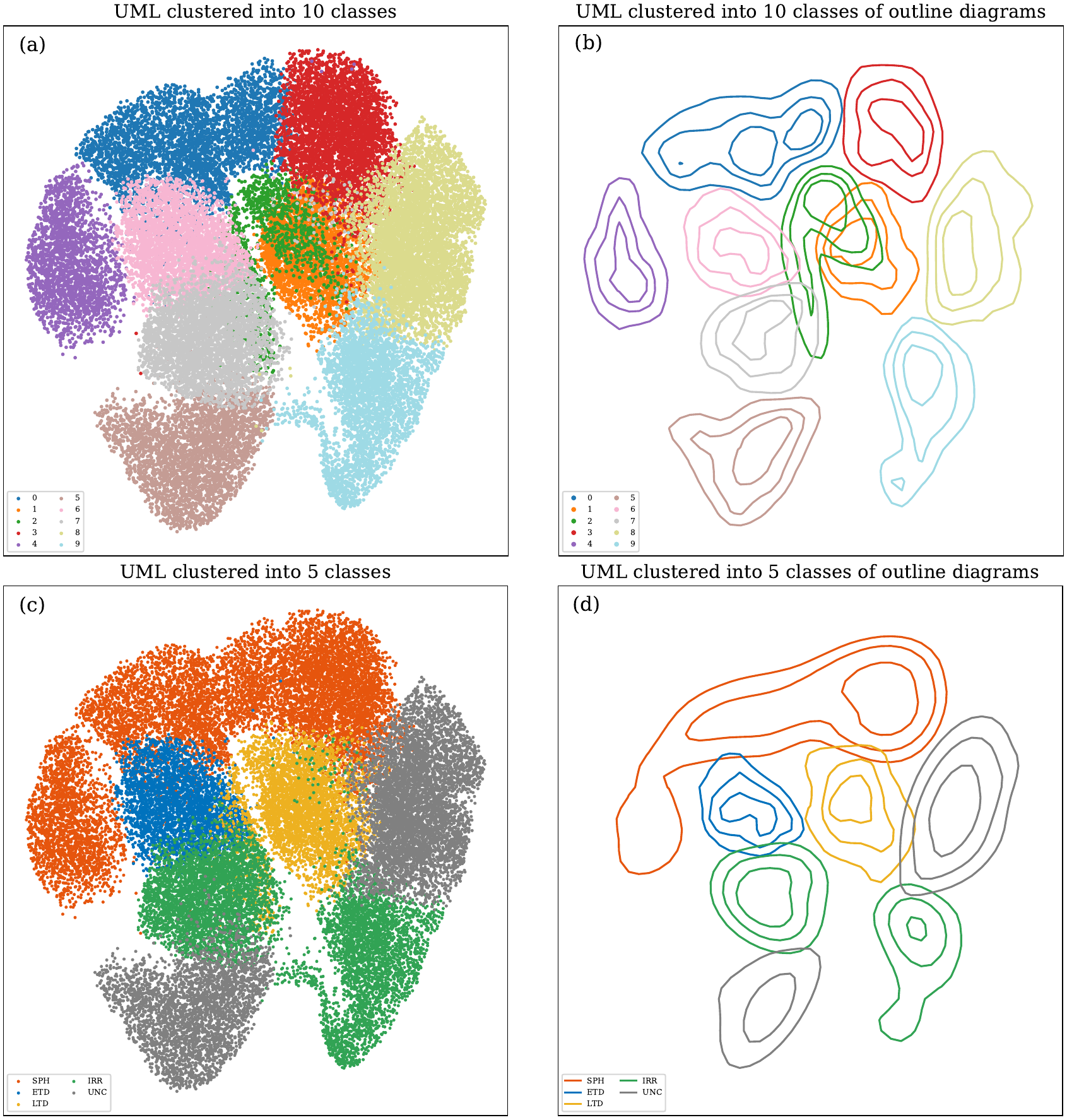}
    \caption{UMAP visualization of the complete classification results. Panels (a) show the two-dimensional mapping results for all 32,922 data points across 10 classes, while panels (b) display contour lines corresponding to the 20\%, 50\%, and 80\% well-classified samples for these 10 classes. Panels (c) present the two-dimensional mapping results for all data points across 5 classes, with panels (b) also showing contour lines for the 20\%, 50\%, and 80\% well-classified samples for these 5 classes.}
    \label{fig:8}
\end{figure*} 

\begin{figure*}
    \includegraphics[width=2\columnwidth]{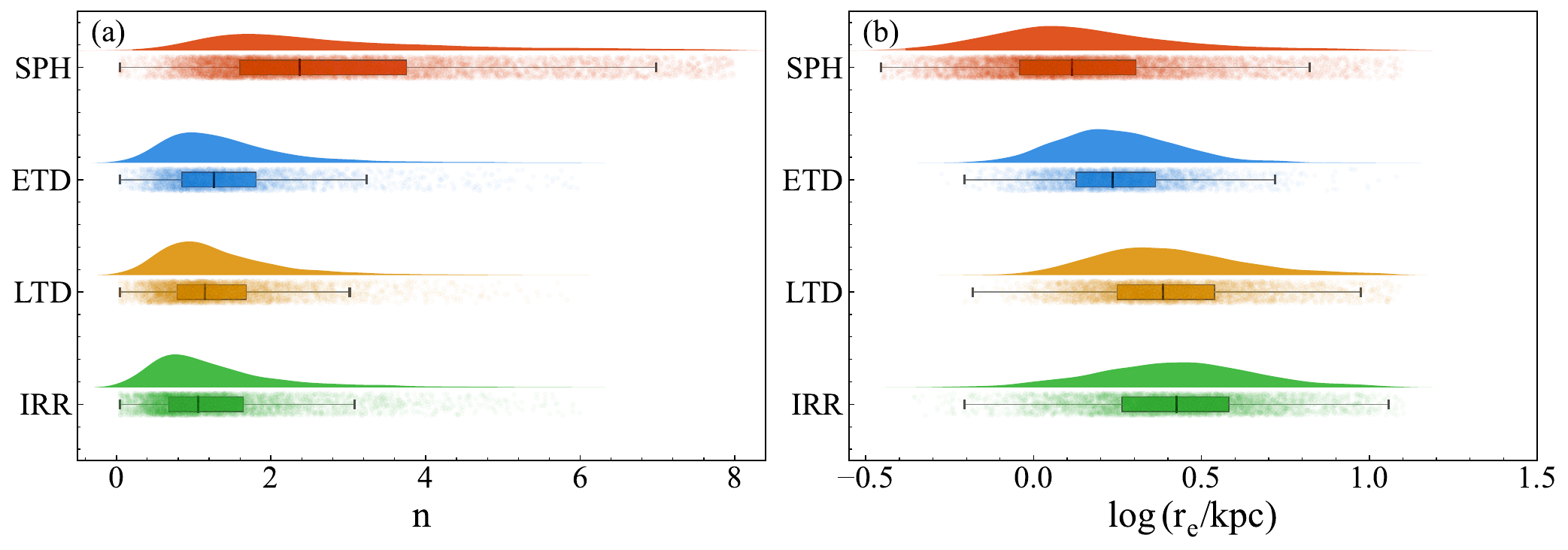}
    \caption{Raincloud plots of S\'{e}rsic index (left panel) and effective radius (right panel) for different types of massive galaxies. The ``cloud'' plots show the distribution of kernel densities in the data; the ``rain'' plots show the distribution of individual values in the dataset. For the box plot, the lower and upper boundaries of the box represent the first and third quartiles, respectively. The black line in the middle of the box represents the median, and the upper and lower whiskers of the box represent the minimum and maximum values. It can be seen that the S\'{e}rsic index of galaxies gradually decreases from SPHs to IRRs, while the effective radius gradually increases.}
    \label{fig:9}
\end{figure*}

\begin{figure*}
	\includegraphics[width=2\columnwidth]{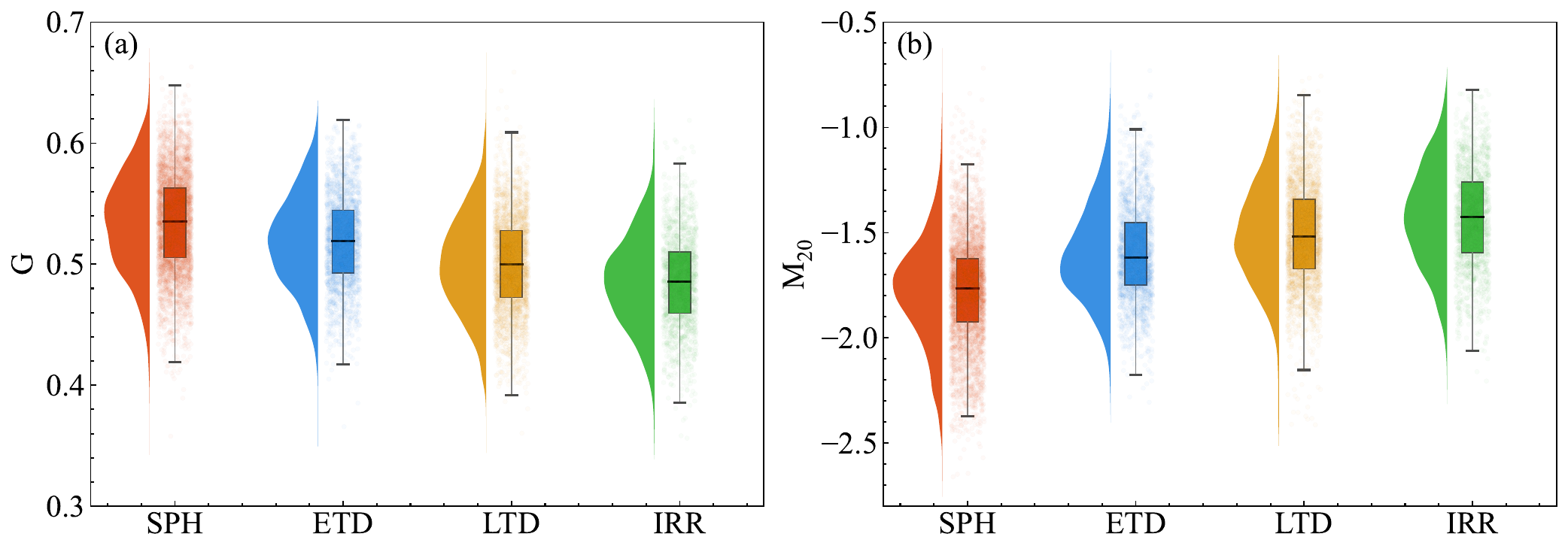}
    \caption{Raincloud plots of G (left panel) and $M_{20}$ (right panel) for different types of massive galaxies, with the black lines in the boxes representing the median values for the different types. It can be seen that from SPHs to IRRs, the G of galaxies gradually decreases, while the $M_{20}$ gradually increases.}
    \label{fig:10}
\end{figure*}

\begin{figure*}

	\includegraphics[width=2\columnwidth]{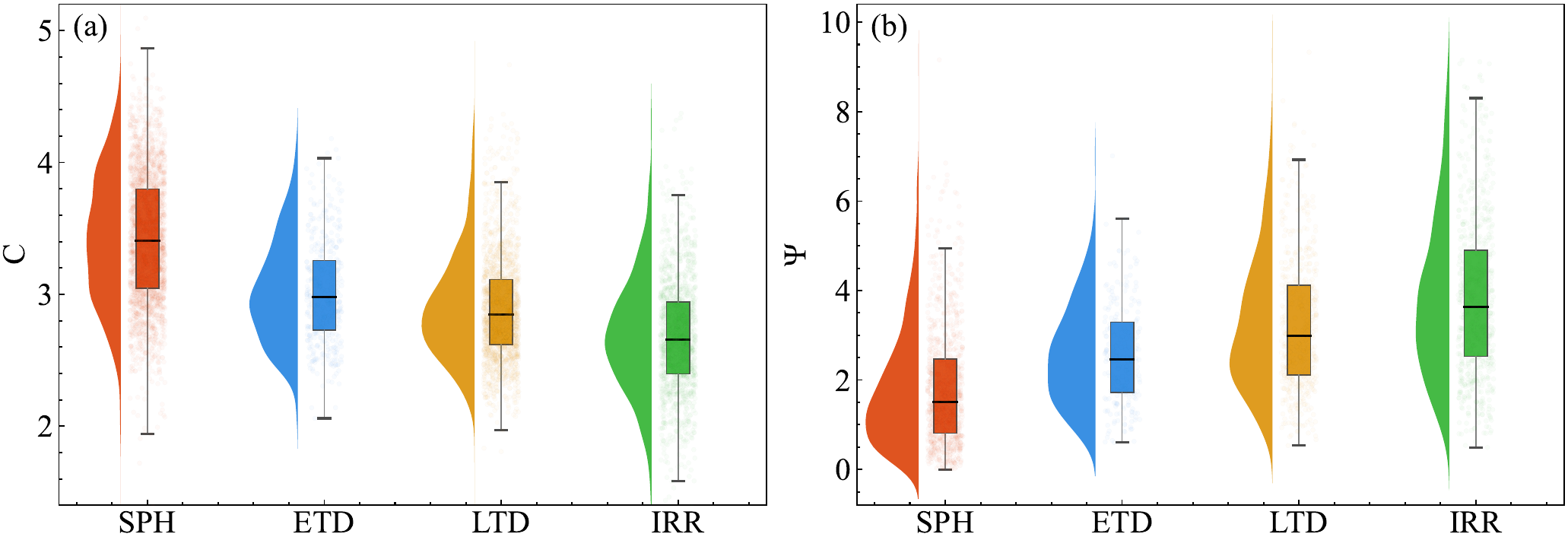}
    \caption{Raincloud plots of C (left panel) and $\Psi$ (right panel) for different types of massive galaxies, with the black lines in the boxes representing the median values for the different types. It can be seen that from SPHs to IRRs, the C of galaxies gradually decreases, while the effective $\Psi$ gradually increases.}
    \label{fig:11}
\end{figure*}
\begin{figure*}
    \includegraphics[width=2\columnwidth]{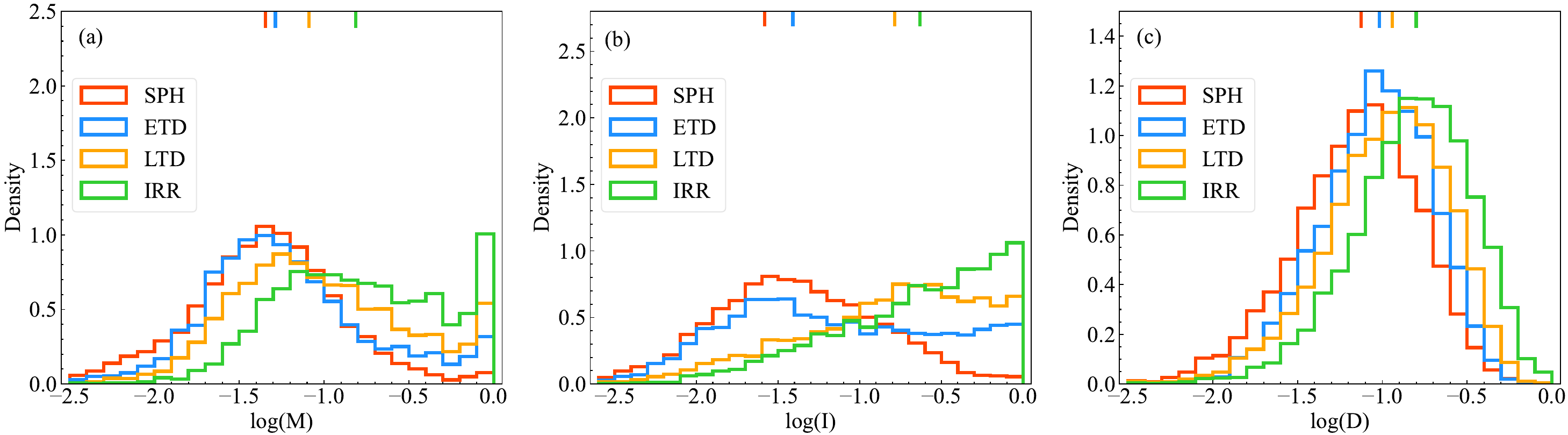}
    \caption{Distributions of M (a panel), I (b panel), and D (c panel) for different types of massive galaxies. The top bars represent the median values of M, I, and D for different types of galaxies. The MIDs of the galaxies are gradually increasing from SPHs to IRRs.}
    \label{fig:12}
\end{figure*}

\section{RESULTS AND DISCUSSION}\label{sec:4}
\subsection{Overall morphological classification results}
By integrating the UML and SML techniques, we succeeded in comprehensively classifying the morphology of 45,288 galaxies in the COSMOS-web field.
As indicated in Table \ref{tab3}, the final classification results contain 14,195 SPHs, 3,986 ETDs, 7,204 LTDs, 8,946 IRRs, and 10,957 UNCs.
Figure \ref{fig:7} shows a classification display of some of the samples, from which it can be seen that different types of galaxies are clearly differentiated morphologically. 
SPHs are more uniformly distributed and exhibit an overall near-circular appearance.
ETDs, on the other hand, are characterized by a prominent central nucleus bulge with an inconspicuous disk. LTDs are marked by pronounced spiral arms and/or a prominent disk structure. IRRs display a variety of irregular appearances, while UNC galaxies are difficult to identify due to their extremely low S/Ns.

In addition, we present the clustering results using the UMAP method, an advanced dimensionality reduction technique particularly well-suited for the visualization of high-dimensional data. UMAP is capable of efficiently mapping complex, high-dimensional data into a low-dimensional space while preserving both local similarities and global structural features of the data \citep{McInnes2018UMAPUM}. As illustrated in Figure \ref{fig:8}, the final UMAP-distributed clustering for all 32,922 galaxies is displayed. The visualization performs exceptionally well in both scenarios: when clustering the galaxies into 10 classes and when dividing them into 5 classes based on visual inspection. Galaxies with similar features are accurately mapped to neighboring positions in the low-dimensional space, while those with different features exhibit clear distributional differences. As evident from the contour plots of sample distributions in panel (d), over 80\% of the samples show no overlap after the UMAP transformation, and the boundaries between clusters are distinctly recognizable. However, the UMAP visualization provides only a qualitative overview. Therefore, we will further validate the classification results using physical properties in the subsequent section.

\subsection{Validation with Galaxy Morphology and Structure Parameters}
Galaxies with varying morphologies often possess distinct physical properties, which are intimately linked to the formation and evolution of galaxies \citep{10.1111/j.1365-2966.2007.12627.x, Gu_2018}. The distribution of galaxy morphology parameters can be used to test the effectiveness of classification results (e.g., \citealt{Zhou_2022, Dai_2023, Yao_2023, Song_2024,Fang2024}). 

Due to the unclear evolution mechanism of low-mass galaxies, we only considered massive galaxies with $M_{*}>10^{9}~M_{\sun}$ for testing. We will exclude the UNC type in the following analysis due to their relatively low S/Ns and the difficulty of measuring their physical properties.

\subsubsection{Parametric Measurements} 
To accurately measure the parameters of galaxies, this study employs the {\tt\string GALAPAGOS} software \citep{Barden2012GALAPAGOSFP, 2022A&A...664A..92H} to conduct parametric morphological analyses on galaxies within the COSMOS-Web field. {\tt\string GALAPAGOS} acts as an interface for both SExtractor \citep{refId0} and {\tt\string GALFIT} \citep{Peng_2002}. By utilizing a single S\'{e}rsic model, the software fits the surface brightness profiles of galaxies to determine S\'{e}rsic index ($n$) and effective radius ($r_e$). To ensure the reliability of our analysis, we followed the setup of
\cite{Song_2024} by excluding galaxies with estimated $r_e$ exceeding 50 pixels. This criterion led to the removal of less than 5\% of the galaxy sample, indicating a negligible impact on our overall results.

In Figure \ref{fig:9}, Panel (a) illustrates the distribution of the S\'{e}rsic indices for the four types of galaxies, with median values for SPHs, ETDs, LTDs, and IRRs being 2.37, 1.26, 1.15, and 1.06, respectively. Generally, elliptical galaxies display larger compactness and typically have S\'{e}rsic indices greater than 2, while disk galaxies have $n$ generally below 2 (e.g., \citealt{fisherSTRUCTURECLASSICALBULGES2008, blantonPhysicalPropertiesEnvironments2009a}), demonstrating a decreasing trend from SPHs to IRRs.
Panel (b) depicts the distribution of the effective radii for the four types of galaxies, with median values for SPHs, ETDs, LTDs, and IRRs at 1.29, 1.74, 2.40, and 2.70 kpc, respectively, indicating a steady increase. Overall, the trends of the S\'{e}rsic index and effective radius across different galaxy types align with our expectations regarding the relationship between galaxy type and structural parameters.

However, while these two parameters are primarily useful for distinguishing between nucleus-dominated and disk-dominated galaxies, they are not as sensitive to other galaxy types, such as barred spirals. To further substantiate the reliability of our classification outcomes, additional parameters will be employed.

\subsubsection{Nonparametric Measurements}
Nonparametric measurements allow us to assess the classification results from several perspectives. Typically, the system employs several real numbers to characterize the overall morphology of galaxies, commonly including the concentration index C (compactness; \citealt{Conselice_2000, Conselice_2003}), the asymmetry index A (asymmetry; \citealt{Conselice_2000, Conselice_2003}), the clumping index S (smoothness), the G (Gini coefficient; \citealt{Lotz_2004,Lotz_2006}), and the second-order moments of the brightest 20\% of the galaxy flux, known as the $M_{20}$ (moments-20; \citealt{Lotz_2004,Lotz_2006}).
Recently, \cite{10.1093/mnras/stt1016} also introduced the MID statistic, which stands for multimode (M), surface density (I), and deviation (D). We use \texttt{statmorph\_csst} \citep{Yao_2023} to compute the CAS, $G$-$M_{20}$, and MID statistics for our sample of galaxies. In addition, we measured the $\Psi$ \citep{Law_2007} and $G_{2}$ \citep{2018MNRAS.477L.101R} parameter as a supplementary metric.Given the widespread use of these parameters in numerous studies, detailed definitions and calculations can be found in the referenced literature.

From the raincloud plot in Figure \ref{fig:10}, 
the Gini coefficient quantifies the inequality in a galaxy's surface brightness distribution, with higher values indicating greater inhomogeneity. The $M_{20}$ index reflects the degree of light concentration in galaxies. A more negative $M_{20}$ indicates a higher concentration of light within the image, but this concentration is not confined to the central region and could be located anywhere in the image (\citealt{Lotz_2004,Lotz_2006}).

We analyze the trends of the Gini and $M_{20}$ parameters for different types of galaxies. Specifically, 
the flux distribution of galaxies exhibits progressively increasing non-uniformity from SPH to IRR types, accompanied by a corresponding gradual decrease in their Gini coefficients, 
with median values of 0.54, 0.52, 0.50, and 0.48, respectively. Conversely, the $M_{20}$ values of galaxies exhibit a gradual increase, with median values of -1.77, -1.62, -1.53, and -1.43, respectively. SPH galaxies typically display the highest Gini coefficients and the lowest $M_{20}$ values, whereas IRR galaxies show the opposite characteristics. The observed trend from IRR to SPH in the figure aligns with our expectations \citep{Zhou_2022,Fang_2023,Dai_2023,Song_2024,Fang2024}.

In Figure \ref{fig:11}, we show the distribution of the C and $\Psi$ parameters for different types of galaxies. The C parameter quantifies the concentration of a galaxy's brightness, with higher values meaning more central concentration (\citealt{Conselice_2000, Conselice_2003}). The $\Psi$ parameter describes the clumpy structures in a galaxy's light distribution (\citealt{Law_2007}). From SPH to IRR, galaxy concentration decreases and clumpy structures become more prominent. Specifically, the median C values for each galaxy type decrease to 3.23, 2.93, 2.82, and 2.75, respectively, while the median $\Psi$ values increase to 1.74, 2.64, 3.45, and 4.04, respectively.
More compact galaxies tend to exhibit larger C values, while more clustered galaxies show higher $\Psi$ values. Our classification outcomes align with the observed patterns of variation \citep{Song_2024,Fang2024}.

The MID parameter can be used to measure the irregularity of galaxy morphology \citep{10.1093/mnras/stt1016}. The larger the MID value, the more irregular the galaxy's morphology. Conversely, more symmetric galaxies tend to have smaller MID values. As illustrated in Figure \ref{fig:12}, the relatively compact and symmetric SPH and ETD galaxies exhibit smaller MID values, whereas the more diffuse and irregular LTD and IRR galaxies show larger MID values. This observation validates the accuracy of our classification \citep{Fang2024}.

\section{Summary} \label{sec:5}
In this work, we update the UML model of the \texttt{USmorph} framework by introducing contrastive learning and integrating it with the SML (GoogLeNet) model to form a closed-loop processing flow. The improved framework consists mainly of the following four steps. (1) The raw images are preprocessed using the CAE and APCT techniques to reduce image noise and enhance rotational invariance. (2) After encoding and extracting features from the preprocessed image data using the pre-trained ConvNeXt and ViT models, contrastive learning is applied to further reduce the dimensionality of image features. (3) Utilizing a clustering model based on Bagging, galaxies are divided into 10 groups. After visual inspection and the removal of sources with inconsistent votes, 73\% (32,922) of the galaxies were successfully classified. (4) We then train the SML model (GoogLeNet) with the labels of galaxies successfully classified to identify the remaining 27\% (12,366) of the sources. Ultimately, this approach classifies 45,288 sources in the COSMOS-Web field into 5 categories: SPHs, ETDs, LTDs, IRRs, and UNCs.

In addition, for further validation of the reliability of the classification, we conducted UMAP visualization tests and relevant physical parameter tests on the results of the machine clustering. 
Different types of galaxies exhibit distinct boundaries in the UMAP visualization. Moreover, the distribution of morphological structural parameters for various galaxy types shows strong consistency with the theory of galaxy evolution. This suggests the reliability of our classification results. Table \ref{tab5} presents a partial catalog of morphological parameters.

We enhance our framework by incorporating contrastive learning, which reduces the discard rate in the voting step and improves the efficiency of visual inspection. By reducing the clustering groups to 10 groups, the visual inspection step takes only a few minutes, the lower discard rate ensures that more labeled galaxies are available after the UML stage. This means there is more valuable information to train our SML model, allowing it to classify rejected galaxies more accurately.

In the future, we aspire to develop more efficient encoding of large models, as well as better dimensionality reduction techniques to go through the accuracy of machine learning models. In addition, we try to apply the popular Agent task with multimodal large
models to the galaxy morphology classification task to further optimize the classification and improve the generalization ability of the model.

\begin{acknowledgements}
This work is supported by the National Science Foundation of China (NSFC, grant No. 12233008), the National Key R\&D Program of China (2023YFA1608100), the Strategic Priority Research Program of the Chinese Academy of Sciences (grant No. XDB0550200), the Cyrus Chun Ying Tang Foundations, and the 111 Project for “Observational and Theoretical Research on Dark Matter and Dark Energy” (B23042). Z.L. acknowledges the support from Hong Kong Innovation and Technology Fund through the Research Talent Hub program  (PiH/022/22GS). The numerical calculations in this paper were done on the computing facilities in the High Performance Computing Platform of Anqing Normal University.

\end{acknowledgements}
\textbf{}

\bibliography{ref}

@ARTICLE{1926ApJ....64..321H,
       author = {{Hubble}, E.~P.},
        title = "{Extragalactic nebulae.}",
      journal = {\apj},
         year = 1926,
        month = dec,
       volume = {64},
        pages = {321-369},
          doi = {10.1086/143018},
       adsurl = {https://ui.adsabs.harvard.edu/abs/1926ApJ....64..321H}
}

@ARTICLE{2024MNRAS.tmp.1656M,
       author = {{Mukundan}, Kavya and {Nair}, Preethi and {Bailin}, Jeremy and {Li}, Wenhao},
        title = "{Automating galaxy morphology classification using K nearest neighbors and non-parametric statistics}",
      journal = {\mnras},
         year = 2024,
        month = jul,
          doi = {10.1093/mnras/stae1684},
       adsurl = {https://ui.adsabs.harvard.edu/abs/2024MNRAS.tmp.1656M}
}

@ARTICLE{2024MNRAS.531.4070D,
       author = {{Desmons}, Alice and {Brough}, Sarah and {Lanusse}, Francois},
        title = "{Detecting galaxy tidal features using self-supervised representation learning}",
      journal = {\mnras},
         year = 2024,
        month = jul,
       volume = {531},
       number = {4},
        pages = {4070-4084},
          doi = {10.1093/mnras/stae1402},
archivePrefix = {arXiv},
       eprint = {2307.04967},
 primaryClass = {astro-ph.GA},
       adsurl = {https://ui.adsabs.harvard.edu/abs/2024MNRAS.531.4070D}
}

@article{Bai_2018,
doi = {10.1088/1674-4527/18/10/118},
url = {https://dx.doi.org/10.1088/1674-4527/18/10/118},
year = {2018},
month = {oct},
publisher = {National Astronomical Observatories, CAS and IOP Publishing Ltd.},
volume = {18},
number = {10},
pages = {118},
author = {Yu Bai and Ji-Feng Liu and Song Wang},
title = {Machine learning classification of Gaia Data Release 2},
journal = {Research in Astronomy and Astrophysics}
}

@article{SCHMIDHUBER201585,
title = {Deep learning in neural networks: An overview},
journal = {Neural Networks},
volume = {61},
pages = {85-117},
year = {2015},
issn = {0893-6080},
doi = {https://doi.org/10.1016/j.neunet.2014.09.003},
url = {https://www.sciencedirect.com/science/article/pii/S0893608014002135},
author = {Jürgen Schmidhuber}
}

@article{1984A&A...132..389P,
  title = {The Typical Interstellar Extinction in the {{Small Magellanic Cloud}}.},
  author = {Prevot, M. L. and Lequeux, J. and Maurice, E. and Prevot, L. and {Rocca-Volmerange}, B.},
  year = {1984},
  journal = {Astronomy \& Astrophysics},
  volume = {132},
  pages = {389--392},
  adsurl = {https://ui.adsabs.harvard.edu/abs/1984A\&A...132..389P}
}

@INPROCEEDINGS{2021AAS...23811902W,
       author = {{Walmsley}, M.},
        title = "{Galaxy Zoo DECaLS: Detailed Visual Morphology Measurements from Volunteers and Bayesian Deep Learning}",
    booktitle = {American Astronomical Society Meeting Abstracts},
         year = 2021,
       series = {American Astronomical Society Meeting Abstracts},
       volume = {238},
        month = jun,
          eid = {119.02},
        pages = {119.02},
       adsurl = {https://ui.adsabs.harvard.edu/abs/2021AAS...23811902W}
}

@ARTICLE{2011MNRAS.410..166L,
       author = {{Lintott}, Chris and {Schawinski}, Kevin and {Bamford}, Steven and {Slosar}, An{\r{a}}{\textthreequarters}e and {Land}, Kate and {Thomas}, Daniel and {Edmondson}, Edd and {Masters}, Karen and {Nichol}, Robert C. and {Raddick}, M. Jordan and {Szalay}, Alex and {Andreescu}, Dan and {Murray}, Phil and {Vandenberg}, Jan},
        title = "{Galaxy Zoo 1: data release of morphological classifications for nearly 900 000 galaxies}",
      journal = {\mnras},
         year = 2011,
        month = jan,
       volume = {410},
       number = {1},
        pages = {166-178},
          doi = {10.1111/j.1365-2966.2010.17432.x},
archivePrefix = {arXiv},
       eprint = {1007.3265},
 primaryClass = {astro-ph.GA},
       adsurl = {https://ui.adsabs.harvard.edu/abs/2011MNRAS.410..166L}
}

@ARTICLE{Tohill_2024,
       author = {{Tohill}, C. and {Bamford}, S.~P. and {Conselice}, C.~J. and {Ferreira}, L. and {Harvey}, T. and {Adams}, N. and {Austin}, D.},
        title = "{A Robust Study of High-redshift Galaxies: Unsupervised Machine Learning for Characterizing Morphology with JWST up to z {\ensuremath{\sim}} 8}",
      journal = {\apj},
         year = 2024,
        month = feb,
       volume = {962},
       number = {2},
          eid = {164},
        pages = {164},
          doi = {10.3847/1538-4357/ad17b8},
archivePrefix = {arXiv},
       eprint = {2306.17225},
 primaryClass = {astro-ph.GA},
       adsurl = {https://ui.adsabs.harvard.edu/abs/2024ApJ...962..164T}
}

@ARTICLE{2023ApJ...954...31C,
       author = {{Casey}, Caitlin M. and {Kartaltepe}, Jeyhan S. and {Drakos}, Nicole E. and {Franco}, Maximilien and {Harish}, Santosh and {Paquereau}, Louise and {Ilbert}, Olivier and {Rose}, Caitlin and {Cox}, Isabella G. and {Nightingale}, James W. and {Robertson}, Brant E. and {Silverman}, John D. and {Koekemoer}, Anton M. and {Massey}, Richard and {McCracken}, Henry Joy and {Rhodes}, Jason and {Akins}, Hollis B. and {Allen}, Natalie and {Amvrosiadis}, Aristeidis and {Arango-Toro}, Rafael C. and {Bagley}, Micaela B. and {Bongiorno}, Angela and {Capak}, Peter L. and {Champagne}, Jaclyn B. and {Chartab}, Nima and {Ch{\'a}vez Ortiz}, {\'O}scar A. and {Chworowsky}, Katherine and {Cooke}, Kevin C. and {Cooper}, Olivia R. and {Darvish}, Behnam and {Ding}, Xuheng and {Faisst}, Andreas L. and {Finkelstein}, Steven L. and {Fujimoto}, Seiji and {Gentile}, Fabrizio and {Gillman}, Steven and {Gould}, Katriona M.~L. and {Gozaliasl}, Ghassem and {Hayward}, Christopher C. and {He}, Qiuhan and {Hemmati}, Shoubaneh and {Hirschmann}, Michaela and {Jahnke}, Knud and {Jin}, Shuowen and {Khostovan}, Ali Ahmad and {Kokorev}, Vasily and {Lambrides}, Erini and {Laigle}, Clotilde and {Larson}, Rebecca L. and {Leung}, Gene C.~K. and {Liu}, Daizhong and {Liaudat}, Tobias and {Long}, Arianna S. and {Magdis}, Georgios and {Mahler}, Guillaume and {Mainieri}, Vincenzo and {Manning}, Sinclaire M. and {Maraston}, Claudia and {Martin}, Crystal L. and {McCleary}, Jacqueline E. and {McKinney}, Jed and {McPartland}, Conor J.~R. and {Mobasher}, Bahram and {Pattnaik}, Rohan and {Renzini}, Alvio and {Rich}, R. Michael and {Sanders}, David B. and {Sattari}, Zahra and {Scognamiglio}, Diana and {Scoville}, Nick and {Sheth}, Kartik and {Shuntov}, Marko and {Sparre}, Martin and {Suzuki}, Tomoko L. and {Talia}, Margherita and {Toft}, Sune and {Trakhtenbrot}, Benny and {Urry}, C. Megan and {Valentino}, Francesco and {Vanderhoof}, Brittany N. and {Vardoulaki}, Eleni and {Weaver}, John R. and {Whitaker}, Katherine E. and {Wilkins}, Stephen M. and {Yang}, Lilan and {Zavala}, Jorge A.},
        title = "{COSMOS-Web: An Overview of the JWST Cosmic Origins Survey}",
      journal = {\apj},
         year = 2023,
        month = sep,
       volume = {954},
       number = {1},
          eid = {31},
        pages = {31},
          doi = {10.3847/1538-4357/acc2bc},
archivePrefix = {arXiv},
       eprint = {2211.07865},
 primaryClass = {astro-ph.GA},
       adsurl = {https://ui.adsabs.harvard.edu/abs/2023ApJ...954...31C}
}

@article{BruzualCharlot,
    author = {Bruzual, G. and Charlot, S.},
    title = "{Stellar population synthesis at the resolution of 2003}",
    journal = {Monthly Notices of the Royal Astronomical Society},
    volume = {344},
    number = {4},
    pages = {1000-1028},
    year = {2003},
    month = {10},
    issn = {0035-8711},
    doi = {10.1046/j.1365-8711.2003.06897.x},
    url = {https://doi.org/10.1046/j.1365-8711.2003.06897.x},
    eprint = {https://academic.oup.com/mnras/article-pdf/344/4/1000/2908334/344-4-1000.pdf},
}

@article{Bates_2019,
    author = {Bates, Dominic J and Tojeiro, Rita and Newman, Jeffrey A and Gonzalez-Perez, Violeta and Comparat, Johan and Schneider, Donald P and Lima, Marcos and Streblyanska, Alina},
    title = {Mass functions, luminosity functions, and completeness measurements from clustering redshifts},
    journal = {Monthly Notices of the Royal Astronomical Society},
    volume = {486},
    number = {3},
    pages = {3059-3077},
    year = {2019},
    month = {04},
    issn = {0035-8711},
    doi = {10.1093/mnras/stz997},
    url = {https://doi.org/10.1093/mnras/stz997},
    eprint = {https://academic.oup.com/mnras/article-pdf/486/3/3059/28531410/stz997.pdf},
}

@ARTICLE{10.1093/mnras/stx2351,
       author = {{Hocking}, Alex and {Geach}, James E. and {Sun}, Yi and {Davey}, Neil},
        title = "{An automatic taxonomy of galaxy morphology using unsupervised machine learning}",
      journal = {\mnras},
         year = 2018,
        month = jan,
       volume = {473},
       number = {1},
        pages = {1108-1129},
          doi = {10.1093/mnras/stx2351},
archivePrefix = {arXiv},
       eprint = {1709.05834},
 primaryClass = {astro-ph.IM},
       adsurl = {https://ui.adsabs.harvard.edu/abs/2018MNRAS.473.1108H}
}

@article{10.1093/mnras/stab734,
    author = {Cheng, Ting-Yun and Huertas-Company, Marc and Conselice, Christopher J and Aragón-Salamanca, Alfonso and Robertson, Brant E and Ramachandra, Nesar},
    title = "{Beyond the hubble sequence – exploring galaxy morphology with unsupervised machine learning}",
    journal = {Monthly Notices of the Royal Astronomical Society},
    volume = {503},
    number = {3},
    pages = {4446-4465},
    year = {2021},
    month = {03},
    issn = {0035-8711},
    doi = {10.1093/mnras/stab734},
    url = {https://doi.org/10.1093/mnras/stab734},
    eprint = {https://academic.oup.com/mnras/article-pdf/503/3/4446/39066819/stab734.pdf},
}

@article{Fang2024,
  author        = {Guanwen Fang and Yao Dai and Zesen Lin and Chichun Zhou and Jie Song and Yizhou Gu and Xiaotong Guo and Anqi Mao and Xu Kong},
  title         = {An efficient unsupervised classification model for galaxy morphology: Voting clustering based on coding from ConvNeXt large model},
  journal       = {Astronomy \& Astrophysics},
  year          = {2024},
  doi           = {10.1051/0004-6361/202451734},
  note          = {Forthcoming article}
}

@ARTICLE{2018MNRAS.477L.101R,
       author = {{Rosa}, R.~R. and {de Carvalho}, R.~R. and {Sautter}, R.~A. and {Barchi}, P.~H. and {Stalder}, D.~H. and {Moura}, T.~C. and {Rembold}, S.~B. and {Morell}, D.~R.~F. and {Ferreira}, N.~C.},
        title = "{Gradient pattern analysis applied to galaxy morphology}",
      journal = {\mnras},
         year = 2018,
        month = jun,
       volume = {477},
       number = {1},
        pages = {L101-L105},
          doi = {10.1093/mnrasl/sly054},
archivePrefix = {arXiv},
       eprint = {1803.10853},
 primaryClass = {astro-ph.GA},
       adsurl = {https://ui.adsabs.harvard.edu/abs/2018MNRAS.477L.101R}
}

@article{Chen2020ImprovedBW,
  title={Improved Baselines with Momentum Contrastive Learning},
  author={Xinlei Chen and Haoqi Fan and Ross B. Girshick and Kaiming He},
  journal={ArXiv},
  year={2020},
  volume={abs/2003.04297},
  url={https://api.semanticscholar.org/CorpusID:212633993}
}

@article{Chen2020ASF,
  title={A Simple Framework for Contrastive Learning of Visual Representations},
  author={Ting Chen and Simon Kornblith and Mohammad Norouzi and Geoffrey E. Hinton},
  journal={ArXiv},
  year={2020},
  volume={abs/2002.05709},
  url={https://api.semanticscholar.org/CorpusID:211096730}
}

@INPROCEEDINGS{10205236,
  author={Woo, Sanghyun and Debnath, Shoubhik and Hu, Ronghang and Chen, Xinlei and Liu, Zhuang and Kweon, In So and Xie, Saining},
  booktitle={2023 IEEE/CVF Conference on Computer Vision and Pattern Recognition (CVPR)}, 
  title={ConvNeXt V2: Co-designing and Scaling ConvNets with Masked Autoencoders}, 
  year={2023},
  volume={},
  number={},
  pages={16133-16142},
  doi={10.1109/CVPR52729.2023.01548}}

@article{Dosovitskiy2020AnII,
  title={An Image is Worth 16x16 Words: Transformers for Image Recognition at Scale},
  author={Alexey Dosovitskiy and Lucas Beyer and Alexander Kolesnikov and Dirk Weissenborn and Xiaohua Zhai and Thomas Unterthiner and Mostafa Dehghani and Matthias Minderer and Georg Heigold and Sylvain Gelly and Jakob Uszkoreit and Neil Houlsby},
  journal={ArXiv},
  year={2020},
  volume={abs/2010.11929},
  url={https://api.semanticscholar.org/CorpusID:225039882}
}

@article{McInnes2018UMAPUM,
  title={UMAP: Uniform Manifold Approximation and Projection for Dimension Reduction},
  author={Leland McInnes and John Healy},
  journal={ArXiv},
  year={2018},
  volume={abs/1802.03426},
  url={https://api.semanticscholar.org/CorpusID:3641284}
}

@article{Law_2007,
doi = {10.1086/510357},
url = {https://dx.doi.org/10.1086/510357},
year = {2007},
month = {feb},
publisher = {},
volume = {656},
number = {1},
pages = {1},
author = {David R. Law and Charles C. Steidel and Dawn K. Erb and Max Pettini and Naveen A. Reddy and Alice E. Shapley and Kurt L. Adelberger and David J. Simenc},
title = {The Physical Nature of Rest-UV Galaxy Morphology during the Peak Epoch of Galaxy Formation},
journal = {The Astrophysical Journal}
}

@ARTICLE{2010A&A...523A..13P,
       author = {{Pozzetti}, L. and {Bolzonella}, M. and {Zucca}, E. and {Zamorani}, G. and {Lilly}, S. and {Renzini}, A. and {Moresco}, M. and {Mignoli}, M. and {Cassata}, P. and {Tasca}, L. and {Lamareille}, F. and {Maier}, C. and {Meneux}, B. and {Halliday}, C. and {Oesch}, P. and {Vergani}, D. and {Caputi}, K. and {Kova{\v{c}}}, K. and {Cimatti}, A. and {Cucciati}, O. and {Iovino}, A. and {Peng}, Y. and {Carollo}, M. and {Contini}, T. and {Kneib}, J. -P. and {Le F{\'e}vre}, O. and {Mainieri}, V. and {Scodeggio}, M. and {Bardelli}, S. and {Bongiorno}, A. and {Coppa}, G. and {de la Torre}, S. and {de Ravel}, L. and {Franzetti}, P. and {Garilli}, B. and {Kampczyk}, P. and {Knobel}, C. and {Le Borgne}, J. -F. and {Le Brun}, V. and {Pell{\`o}}, R. and {Perez Montero}, E. and {Ricciardelli}, E. and {Silverman}, J.~D. and {Tanaka}, M. and {Tresse}, L. and {Abbas}, U. and {Bottini}, D. and {Cappi}, A. and {Guzzo}, L. and {Koekemoer}, A.~M. and {Leauthaud}, A. and {Maccagni}, D. and {Marinoni}, C. and {McCracken}, H.~J. and {Memeo}, P. and {Porciani}, C. and {Scaramella}, R. and {Scarlata}, C. and {Scoville}, N.},
        title = "{zCOSMOS - 10k-bright spectroscopic sample. The bimodality in the galaxy stellar mass function: exploring its evolution with redshift}",
      journal = {\aap},
         year = 2010,
        month = nov,
       volume = {523},
          eid = {A13},
        pages = {A13},
          doi = {10.1051/0004-6361/200913020},
archivePrefix = {arXiv},
       eprint = {0907.5416},
 primaryClass = {astro-ph.CO},
       adsurl = {https://ui.adsabs.harvard.edu/abs/2010A&A...523A..13P}
}

@article{Weaver_2022,
  author = {{Weaver}, J.~R. and {Kauffmann}, O.~B. and {Ilbert}, O. and {McCracken}, H.~J. and {Moneti}, A. and {Toft}, S. and {Brammer}, G. and {Shuntov}, M. and {Davidzon}, I. and {Hsieh}, B.~C. and {Laigle}, C. and {Anastasiou}, A. and {Jespersen}, C.~K. and {Vinther}, J. and {Capak}, P. and {Casey}, C.~M. and {McPartland}, C.~J.~R. and {Milvang-Jensen}, B. and {Mobasher}, B. and {Sanders}, D.~B. and {Zalesky}, L. and {Arnouts}, S. and {Aussel}, H. and {Dunlop}, J.~S. and {Faisst}, A. and {Franx}, M. and {Furtak}, L.~J. and {Fynbo}, J.~P.~U. and {Gould}, K.~M.~L. and {Greve}, T.~R. and {Gwyn}, S. and {Kartaltepe}, J.~S. and {Kashino}, D. and {Koekemoer}, A.~M. and {Kokorev}, V. and {Le Fèvre}, O. and {Lilly}, S. and {Masters}, D. and {Magdis}, G. and {Mehta}, V. and {Peng}, Y. and {Riechers}, D.~A. and {Salvato}, M. and {Sawicki}, M. and {Scarlata}, C. and {Scoville}, N. and {Shirley}, R. and {Silverman}, J.~D. and {Sneppen}, A. and {Smolčić}, V. and {Steinhardt}, C. and {Stern}, D. and {Tanaka}, M. and {Taniguchi}, Y. and {Teplitz}, H.~I. and {Vaccari}, M. and {Wang}, W.-H. and {Zamorani}, G.},
  title = {COSMOS2020: A Panchromatic View of the Universe to z ∼ 10 from Two Complementary Catalogs},
  journal = {The Astrophysical Journal Supplement Series},
  year = {2022},
  month = {jan},
  volume = {258},
  number = {1},
  pages = {11},
  doi = {10.3847/1538-4365/ac3078},
  url = {https://dx.doi.org/10.3847/1538-4365/ac3078}
}

@misc{liu2022networkspixelsembeddingmethod,
      title={Networks with pixels embedding: a method to improve noise resistance in images classification}, 
      author={Yang Liu and Hai-Long Tu and Chi-Chun Zhou and Yi Liu and Fu-Lin Zhang},
      year={2022},
      eprint={2005.11679},
      archivePrefix={arXiv},
      primaryClass={cs.CV},
      url={https://arxiv.org/abs/2005.11679}
}

@article{ refId0,
	author = {{Bertin, E.} and {Arnouts, S.}},
	title = {SExtractor: Software for source extraction},
	DOI= "10.1051/aas:1996164",
	url= "https://doi.org/10.1051/aas:1996164",
	journal = {Astron. Astrophys. Suppl. Ser.},
	year = 1996,
	volume = 117,
	number = 2,
	pages = "393-404",
}

@ARTICLE{1983ApJ...266..713O,
       author = {{Oke}, J.~B. and {Gunn}, J.~E.},
        title = "{Secondary standard stars for absolute spectrophotometry.}",
      journal = {\apj},
         year = 1983,
        month = mar,
       volume = {266},
        pages = {713-717},
          doi = {10.1086/160817},
       adsurl = {https://ui.adsabs.harvard.edu/abs/1983ApJ...266..713O}
}

@article{Chabrier_2003,
doi = {10.1086/376392},
url = {https://dx.doi.org/10.1086/376392},
year = {2003},
month = {jul},
publisher = {The University of Chicago Press},
volume = {115},
number = {809},
pages = {763},
author = {Gilles Chabrier},
title = {Galactic Stellar and Substellar Initial Mass Function1},
journal = {Publications of the Astronomical Society of the Pacific},
}

@article{Rieke_2023,
doi = {10.1088/1538-3873/acac53},
url = {https://dx.doi.org/10.1088/1538-3873/acac53},
year = {2023},
month = {feb},
publisher = {The Astronomical Society of the Pacific},
volume = {135},
number = {1044},
pages = {028001},
author = {Marcia J. Rieke and Douglas M. Kelly and Karl Misselt and John Stansberry and Martha Boyer and Thomas Beatty and Eiichi Egami and Michael Florian and Thomas P. Greene and Kevin Hainline and Jarron Leisenring and Thomas Roellig and Everett Schlawin and Fengwu Sun and Lee Tinnin and Christina C. Williams and Christopher N. A. Willmer and Debra Wilson and Charles R. Clark and Scott Rohrbach and Brian Brooks and Alicia Canipe and Matteo Correnti and Audrey DiFelice and Mario Gennaro and Julian Girard and George Hartig and Bryan Hilbert and Anton M. Koekemoer and Nikolay K. Nikolov and Norbert Pirzkal and Armin Rest and Massimo Robberto and Ben Sunnquist and Randal Telfer and Chi Rai Wu and Malcolm Ferry and Dan Lewis and Stefi Baum and Charles Beichman and René Doyon and Alan Dressler and Daniel J. Eisenstein and Laura Ferrarese and Klaus Hodapp and Scott Horner and Daniel T. Jaffe and Doug Johnstone and John Krist and Peter Martin and Donald W. McCarthy and Michael Meyer and George H. Rieke and John Trauger and Erick T. Young},
title = {Performance of NIRCam on JWST in Flight},
journal = {Publications of the Astronomical Society of the Pacific}
}

@INPROCEEDINGS{2022SPIE12180E..3PW,
       author = {{Wright}, Raymond H. and {Sabatke}, Derek and {Telfer}, Randal},
        title = "{James Webb Space Telescope MIRI shear pupil analysis}",
    booktitle = {Space Telescopes and Instrumentation 2022: Optical, Infrared, and Millimeter Wave},
         year = 2022,
       editor = {{Coyle}, Laura E. and {Matsuura}, Shuji and {Perrin}, Marshall D.},
       series = {Society of Photo-Optical Instrumentation Engineers (SPIE) Conference Series},
       volume = {12180},
        month = aug,
          eid = {121803P},
        pages = {121803P},
          doi = {10.1117/12.2632087},
       adsurl = {https://ui.adsabs.harvard.edu/abs/2022SPIE12180E..3PW}
}

@article{fisherSTRUCTURECLASSICALBULGES2008,
  title = {{{THE STRUCTURE OF CLASSICAL BULGES AND PSEUDOBULGES}}: {{THE LINK BETWEEN PSEUDOBULGES AND S\'ERSIC INDEX}}},
  shorttitle = {{{THE STRUCTURE OF CLASSICAL BULGES AND PSEUDOBULGES}}},
  author = {Fisher, David B. and Drory, Niv},
  year = {2008},
  journal = {The Astronomical Journal},
  volume = {136},
  number = {2},
  pages = {773--839},
  doi = {10.1088/0004-6256/136/2/773},
  urldate = {2023-10-07},
  langid = {english}
}

@ARTICLE{blantonPhysicalPropertiesEnvironments2009a,
       author = {{Blanton}, Michael R. and {Moustakas}, John},
        title = "{Physical Properties and Environments of Nearby Galaxies}",
      journal = {\araa},
         year = 2009,
        month = sep,
       volume = {47},
       number = {1},
        pages = {159-210},
          doi = {10.1146/annurev-astro-082708-101734},
archivePrefix = {arXiv},
       eprint = {0908.3017},
 primaryClass = {astro-ph.GA},
       adsurl = {https://ui.adsabs.harvard.edu/abs/2009ARA&A..47..159B}
}

@article{10.1093/mnras/stac562,
  author    = {Wilde, Joshua and Serjeant, Stephen and Bromley, Jane M and Dickinson, Hugh and Koopmans, Léon V E and Metcalf, R Benton},
  title     = {Detecting gravitational lenses using machine learning: exploring interpretability and sensitivity to rare lensing configurations},
  journal   = {Monthly Notices of the Royal Astronomical Society},
  volume    = {512},
  number    = {3},
  pages     = {3464-3479},
  year      = {2022},
  month     = {02},
  issn      = {0035-8711},
  doi       = {10.1093/mnras/stac562},
  url       = {https://doi.org/10.1093/mnras/stac562},
  eprint    = {https://academic.oup.com/mnras/article-pdf/512/3/3464/43249501/stac562.pdf}
}

@article{10.1093/mnras/stab1677,
    author = {Ćiprijanović, A and Kafkes, D and Downey, K and Jenkins, S and Perdue, G N and Madireddy, S and Johnston, T and Snyder, G F and Nord, B},
    title = {DeepMerge – II. Building robust deep learning algorithms for merging galaxy identification across domains},
    journal = {Monthly Notices of the Royal Astronomical Society},
    volume = {506},
    number = {1},
    pages = {677-691},
    year = {2021},
    month = {06},
    issn = {0035-8711},
    doi = {10.1093/mnras/stab1677},
    url = {https://doi.org/10.1093/mnras/stab1677},
    eprint = {https://academic.oup.com/mnras/article-pdf/506/1/677/38890775/stab1677.pdf},
}

@article{Zhang_2024,
doi = {10.1088/1674-4527/ad6fe6},
url = {https://dx.doi.org/10.1088/1674-4527/ad6fe6},
year = {2024},
month = {sep},
publisher = {National Astromonical Observatories, CAS and IOP Publishing},
volume = {24},
number = {9},
pages = {095012},
author = {Shiliang Zhang and Guanwen Fang and Jie Song and Ran Li and Yizhou Gu and Zesen Lin and Chichun Zhou and Yao Dai and Xu Kong},
title = {Preparation for CSST: Star-galaxy Classification using a Rotationally Invariant Supervised Machine Learning Method},
journal = {Research in Astronomy and Astrophysics}
}

@ARTICLE{1958MeLuS.136....1H,
       author = {{Holmberg}, E.},
        title = "{A photographic photometry of extragalactic nebulae.}",
      journal = {Meddelanden fran Lunds Astronomiska Observatorium Serie II},
         year = 1958,
        month = jan,
       volume = {136},
        pages = {1},
       adsurl = {https://ui.adsabs.harvard.edu/abs/1958MeLuS.136....1H}
}

@ARTICLE{1980ApJ...236..351D,
       author = {{Dressler}, A.},
        title = "{Galaxy morphology in rich clusters: implications for the formation and evolution of galaxies.}",
      journal = {\apj},
         year = 1980,
        month = mar,
       volume = {236},
        pages = {351-365},
          doi = {10.1086/157753},
       adsurl = {https://ui.adsabs.harvard.edu/abs/1980ApJ...236..351D}
}

@article{10.1093/mnras/sts682,
    author = {Conselice, Christopher J. and Mortlock, Alice and Bluck, Asa F. L. and Grützbauch, Ruth and Duncan, Kenneth},
    title = {Gas accretion as a dominant formation mode in massive galaxies from the GOODS NICMOS Survey},
    journal = {Monthly Notices of the Royal Astronomical Society},
    volume = {430},
    number = {2},
    pages = {1051-1060},
    year = {2013},
    month = {02},
    issn = {0035-8711},
    doi = {10.1093/mnras/sts682},
    url = {https://doi.org/10.1093/mnras/sts682},
    eprint = {https://academic.oup.com/mnras/article-pdf/430/2/1051/9378804/sts682.pdf},
}

@article{kauffmannEnvironmentalDependenceRelations2004,
  title = {The Environmental Dependence of the Relations between Stellar Mass, Structure, Star Formation and Nuclear Activity in Galaxies: {{Galaxy}} Structure, Star Formation and Nuclear Activity},
  shorttitle = {The Environmental Dependence of the Relations between Stellar Mass, Structure, Star Formation and Nuclear Activity in Galaxies},
  author = {Kauffmann, Guinevere and White, Simon D. M. and Heckman, Timothy M. and M{\'e}nard, Brice and Brinchmann, Jarle and Charlot, St{\'e}phane and Tremonti, Christy and Brinkmann, Jon},
  year = {2004},
  journal = {Monthly Notices of the Royal Astronomical Society},
  volume = {353},
  number = {3},
  pages = {713--731},
  doi = {10.1111/j.1365-2966.2004.08117.x},
  urldate = {2022-08-26},
  langid = {english}
}

@article{omandConnectionGalaxyStructure2014,
  title = {The Connection between Galaxy Structure and Quenching Efficiency},
  author = {Omand, Conor M. B. and Balogh, Michael L. and Poggianti, Bianca M.},
  year = {2014},
  journal = {Monthly Notices of the Royal Astronomical Society},
  volume = {440},
  number = {1},
  pages = {843--858},
  doi = {10.1093/mnras/stu331},
  urldate = {2023-07-11},
  langid = {english}
}

@article{schawinskiGreenValleyRed2014,
  title = {The Green Valley Is a Red Herring: {{Galaxy Zoo}} Reveals Two Evolutionary Pathways towards Quenching of Star Formation in Early- and Late-Type Galaxies\ding{72}},
  shorttitle = {The Green Valley Is a Red Herring},
  author = {Schawinski, Kevin and Urry, C. Megan and Simmons, Brooke D. and Fortson, Lucy and Kaviraj, Sugata and Keel, William C. and Lintott, Chris J. and Masters, Karen L. and Nichol, Robert C. and Sarzi, Marc and Skibba, Ramin and Treister, Ezequiel and Willett, Kyle W. and Wong, O. Ivy and Yi, Sukyoung K.},
  year = {2014},
  journal = {Monthly Notices of the Royal Astronomical Society},
  volume = {440},
  number = {1},
  pages = {889--907},
  doi = {10.1093/mnras/stu327},
  urldate = {2022-08-26},
  langid = {english}
}

@article{guMorphologicalEvolutionAGN2018,
  title = {The {{Morphological Evolution}}, {{AGN Fractions}}, {{Dust Content}}, {{Environments}}, and {{Downsizing}} of {{Massive Green Valley Galaxies}} at 0.5 {$<$} {\emph{z}} {$<$} 2.5 in {{3D-}} {{{\emph{HST}}}} /{{CANDELS}}},
  author = {Gu, Yizhou and Fang, Guanwen and Yuan, Qirong and Cai, Zhenyi and Wang, Tao},
  year = {2018},
  journal = {The Astrophysical Journal},
  volume = {855},
  number = {1},
  pages = {10},
  doi = {10.3847/1538-4357/aaad0b},
  urldate = {2023-02-20},
  langid = {english}
}

@article{lianouDustPropertiesStar2019,
  title = {Dust Properties and Star Formation of Approximately a Thousand Local Galaxies},
  author = {Lianou, S. and Barmby, P. and Mosenkov, A. A. and Lehnert, M. and Karczewski, O.},
  year = {2019},
  journal = {Astronomy \& Astrophysics},
  volume = {631},
  pages = {A38},
  doi = {10.1051/0004-6361/201834553},
  urldate = {2023-07-11},
  langid = {english}
}

@article{kawinwanichakijEffectLocalEnvironment2017,
  title = {Effect of {{Local Environment}} and {{Stellar Mass}} on {{Galaxy Quenching}} and {{Morphology}} at 0.5 {$<$} z {$<$} 2.0},
  author = {Kawinwanichakij, Lalitwadee and Papovich, Casey and Quadri, Ryan F. and Glazebrook, Karl and Kacprzak, Glenn G. and Allen, Rebecca J. and Bell, Eric F. and Croton, Darren J. and Dekel, Avishai and Ferguson, Henry C. and Forrest, Ben and Grogin, Norman A. and Guo, Yicheng and Kocevski, Dale D. and Koekemoer, Anton M. and Labb{\'e}, Ivo and Lucas, Ray A. and Nanayakkara, Themiya and Spitler, Lee R. and Straatman, Caroline M. S. and Tran, Kim-Vy H. and Tomczak, Adam and van Dokkum, Pieter},
  year = {2017},
  journal = {The Astrophysical Journal},
  volume = {847},
  number = {2},
  pages = {134},
  doi = {10.3847/1538-4357/aa8b75},
  urldate = {2021-12-06},
  langid = {english}
}

@article{Bergh1976ANC,
  title={A new classification system for galaxies.},
  author={Sidney van den Bergh},
  journal={The Astrophysical Journal},
  year={1976},
  volume={206},
  pages={883-887},
  url={https://api.semanticscholar.org/CorpusID:122319125}
}

@ARTICLE{1959HDP....53.....F,
       author = {{Fl{\"u}gge}, S.},
        title = "{Astrophysik IV: Sternsysteme / Astrophysics IV: Stellar Systems}",
      journal = {Handbuch der Physik},
         year = 1959,
        month = jan,
       volume = {11},
          doi = {10.1007/978-3-642-45932-0},
       adsurl = {https://ui.adsabs.harvard.edu/abs/1959HDP....53.....F}
}

@ARTICLE{1996ApJS..107....1A,
       author = {{Abraham}, Roberto G. and {van den Bergh}, Sidney and {Glazebrook}, Karl and {Ellis}, Richard S. and {Santiago}, Basilio X. and {Surma}, Peter and {Griffiths}, Richard E.},
        title = "{The Morphologies of Distant Galaxies. II. Classifications from the Hubble Space Telescope Medium Deep Survey}",
      journal = {\apjs},
         year = 1996,
        month = nov,
       volume = {107},
        pages = {1},
          doi = {10.1086/192352},
       adsurl = {https://ui.adsabs.harvard.edu/abs/1996ApJS..107....1A}
}

@article{Ferreira_2020,
doi = {10.3847/1538-4357/ab8f9b},
url = {https://dx.doi.org/10.3847/1538-4357/ab8f9b},
year = {2020},
month = {jun},
publisher = {The American Astronomical Society},
volume = {895},
number = {2},
pages = {115},
author = {Leonardo Ferreira and Christopher J. Conselice and Kenneth Duncan and Ting-Yun Cheng and Alex Griffiths and Amy Whitney},
title = {Galaxy Merger Rates up to z∼3 Using a Bayesian Deep Learning Model: A Major-merger Classifier Using IllustrisTNG Simulation Data},
journal = {The Astrophysical Journal}
}

@INPROCEEDINGS{8899285,
  author={Yao, Xiwen and Feng, Xiaoxu and Cheng, Gong and Han, Junwei and Guo, Lei},
  booktitle={IGARSS 2019 - 2019 IEEE International Geoscience and Remote Sensing Symposium}, 
  title={Rotation-Invariant Latent Semantic Representation Learning for Object Detection in VHR Optical Remote Sensing Images}, 
  year={2019},
  volume={},
  number={},
  pages={1382-1385},
  doi={10.1109/IGARSS.2019.8899285}}

@ARTICLE{2002AJ....123..485S,
       author = {{Stoughton}, Chris and {Lupton}, Robert H. and {Bernardi}, Mariangela and {Blanton}, Michael R. and {Burles}, Scott and {Castander}, Francisco J. and {Connolly}, A.~J. and {Eisenstein}, Daniel J. and {Frieman}, Joshua A. and {Hennessy}, G.~S. and {Hindsley}, Robert B. and {Ivezi{\'c}}, {\v{Z}}eljko and {Kent}, Stephen and {Kunszt}, Peter Z. and {Lee}, Brian C. and {Meiksin}, Avery and {Munn}, Jeffrey A. and {Newberg}, Heidi Jo and {Nichol}, R.~C. and {Nicinski}, Tom and {Pier}, Jeffrey R. and {Richards}, Gordon T. and {Richmond}, Michael W. and {Schlegel}, David J. and {Smith}, J. Allyn and {Strauss}, Michael A. and {SubbaRao}, Mark and {Szalay}, Alexander S. and {Thakar}, Aniruddha R. and {Tucker}, Douglas L. and {Vanden Berk}, Daniel E. and {Yanny}, Brian and {Adelman}, Jennifer K. and {Anderson}, John E., Jr. and {Anderson}, Scott F. and {Annis}, James and {Bahcall}, Neta A. and {Bakken}, J.~A. and {Bartelmann}, Matthias and {Bastian}, Steven and {Bauer}, Amanda and {Berman}, Eileen and {B{\"o}hringer}, Hans and {Boroski}, William N. and {Bracker}, Steve and {Briegel}, Charlie and {Briggs}, John W. and {Brinkmann}, J. and {Brunner}, Robert and {Carey}, Larry and {Carr}, Michael A. and {Chen}, Bing and {Christian}, Damian and {Colestock}, Patrick L. and {Crocker}, J.~H. and {Csabai}, Istv{\'a}n and {Czarapata}, Paul C. and {Dalcanton}, Julianne and {Davidsen}, Arthur F. and {Davis}, John Eric and {Dehnen}, Walter and {Dodelson}, Scott and {Doi}, Mamoru and {Dombeck}, Tom and {Donahue}, Megan and {Ellman}, Nancy and {Elms}, Brian R. and {Evans}, Michael L. and {Eyer}, Laurent and {Fan}, Xiaohui and {Federwitz}, Glenn R. and {Friedman}, Scott and {Fukugita}, Masataka and {Gal}, Roy and {Gillespie}, Bruce and {Glazebrook}, Karl and {Gray}, Jim and {Grebel}, Eva K. and {Greenawalt}, Bruce and {Greene}, Gretchen and {Gunn}, James E. and {de Haas}, Ernst and {Haiman}, Zolt{\'a}n and {Haldeman}, Merle and {Hall}, Patrick B. and {Hamabe}, Masaru and {Hansen}, Brad and {Harris}, Frederick H. and {Harris}, Hugh and {Harvanek}, Michael and {Hawley}, Suzanne L. and {Hayes}, J.~J.~E. and {Heckman}, Timothy M. and {Helmi}, Amina and {Henden}, Arne and {Hogan}, Craig J. and {Hogg}, David W. and {Holmgren}, Donald J. and {Holtzman}, Jon and {Huang}, Chih-Hao and {Hull}, Charles and {Ichikawa}, Shin-Ichi and {Ichikawa}, Takashi and {Johnston}, David E. and {Kauffmann}, Guinevere and {Kim}, Rita S.~J. and {Kimball}, Tim and {Kinney}, E. and {Klaene}, Mark and {Kleinman}, S.~J. and {Klypin}, Anatoly and {Knapp}, G.~R. and {Korienek}, John and {Krolik}, Julian and {Kron}, Richard G. and {Krzesi{\'n}ski}, Jurek and {Lamb}, D.~Q. and {Leger}, R. French and {Limmongkol}, Siriluk and {Lindenmeyer}, Carl and {Long}, Daniel C. and {Loomis}, Craig and {Loveday}, Jon and {MacKinnon}, Bryan and {Mannery}, Edward J. and {Mantsch}, P.~M. and {Margon}, Bruce and {McGehee}, Peregrine and {McKay}, Timothy A. and {McLean}, Brian and {Menou}, Kristen and {Merelli}, Aronne and {Mo}, H.~J. and {Monet}, David G. and {Nakamura}, Osamu and {Narayanan}, Vijay K. and {Nash}, Thomas and {Neilsen}, Eric H., Jr. and {Newman}, Peter R. and {Nitta}, Atsuko and {Odenkirchen}, Michael and {Okada}, Norio and {Okamura}, Sadanori and {Ostriker}, Jeremiah P. and {Owen}, Russell and {Pauls}, A. George and {Peoples}, John and {Peterson}, R.~S. and {Petravick}, Donald and {Pope}, Adrian and {Pordes}, Ruth and {Postman}, Marc and {Prosapio}, Angela and {Quinn}, Thomas R. and {Rechenmacher}, Ron and {Rivetta}, Claudio H. and {Rix}, Hans-Walter and {Rockosi}, Constance M. and {Rosner}, Robert and {Ruthmansdorfer}, Kurt and {Sandford}, Dale and {Schneider}, Donald P. and {Scranton}, Ryan and {Sekiguchi}, Maki and {Sergey}, Gary and {Sheth}, Ravi and {Shimasaku}, Kazuhiro and {Smee}, Stephen and {Snedden}, Stephanie A. and {Stebbins}, Albert and {Stubbs}, Christopher and {Szapudi}, Istv{\'a}n and {Szkody}, Paula and {Szokoly}, Gyula P. and {Tabachnik}, Serge and {Tsvetanov}, Zlatan and {Uomoto}, Alan and {Vogeley}, Michael S. and {Voges}, Wolfgang and {Waddell}, Patrick and {Walterbos}, Ren{\'e} and {Wang}, Shu-i. and {Watanabe}, Masaru and {Weinberg}, David H. and {White}, Richard L. and {White}, Simon D.~M. and {Wilhite}, Brian and {Wolfe}, David and {Yasuda}, Naoki and {York}, Donald G. and {Zehavi}, Idit and {Zheng}, Wei},
        title = "{Sloan Digital Sky Survey: Early Data Release}",
      journal = {\aj},
         year = 2002,
        month = jan,
       volume = {123},
       number = {1},
        pages = {485-548},
          doi = {10.1086/324741},
       adsurl = {https://ui.adsabs.harvard.edu/abs/2002AJ....123..485S}
}

@ARTICLE{2006ApJ...652..963R,
       author = {{Ravindranath}, Swara and {Giavalisco}, Mauro and {Ferguson}, Henry C. and {Conselice}, Christopher and {Katz}, Neal and {Weinberg}, Martin and {Lotz}, Jennifer and {Dickinson}, Mark and {Fall}, S. Michael and {Mobasher}, Bahram and {Papovich}, Casey},
        title = "{The Morphological Diversities among Star-forming Galaxies at High Redshifts in the Great Observatories Origins Deep Survey}",
      journal = {\apj},
         year = 2006,
        month = dec,
       volume = {652},
       number = {2},
        pages = {963-980},
          doi = {10.1086/507016},
archivePrefix = {arXiv},
       eprint = {astro-ph/0606696},
 primaryClass = {astro-ph},
       adsurl = {https://ui.adsabs.harvard.edu/abs/2006ApJ...652..963R}
}

@article{Scoville_2007,
   title={The Cosmic Evolution Survey (COSMOS): Overview},
   volume={172},
   ISSN={1538-4365},
   url={http://dx.doi.org/10.1086/516585},
   DOI={10.1086/516585},
   number={1},
   journal={The Astrophysical Journal Supplement Series},
   publisher={American Astronomical Society},
   author={Scoville, N. and Aussel, H. and Brusa, M. and Capak, P. and Carollo, C. M. and Elvis, M. and Giavalisco, M. and Guzzo, L. and Hasinger, G. and Impey, C. and Kneib, J.‐P. and LeFevre, O. and Lilly, S. J. and Mobasher, B. and Renzini, A. and Rich, R. M. and Sanders, D. B. and Schinnerer, E. and Schminovich, D. and Shopbell, P. and Taniguchi, Y. and Tyson, N. D.},
   year={2007},
   month=sep, pages={1–8} }

@article{Zhou_2022,
   title={Automatic Morphological Classification of Galaxies: Convolutional Autoencoder and Bagging-based Multiclustering Model},
   volume={163},
   ISSN={1538-3881},
   url={http://dx.doi.org/10.3847/1538-3881/ac4245},
   DOI={10.3847/1538-3881/ac4245},
   number={2},
   journal={The Astronomical Journal},
   publisher={American Astronomical Society},
   author={Zhou, ChiChun and Gu, Yizhou and Fang, Guanwen and Lin, Zesen},
   year={2022},
   month=jan, pages={86} }

@misc{caron2021unsupervised,
      title={Unsupervised Learning of Visual Features by Contrasting Cluster Assignments}, 
      author={Mathilde Caron and Ishan Misra and Julien Mairal and Priya Goyal and Piotr Bojanowski and Armand Joulin},
      year={2021},
      eprint={2006.09882},
      archivePrefix={arXiv},
      primaryClass={cs.CV}
}

@article{Grill2020BootstrapYO,
  title={Bootstrap Your Own Latent: A New Approach to Self-Supervised Learning},
  author={Jean-Bastien Grill and Florian Strub and Florent Altch'e and Corentin Tallec and Pierre H. Richemond and Elena Buchatskaya and Carl Doersch and Bernardo {\'A}vila Pires and Zhaohan Daniel Guo and Mohammad Gheshlaghi Azar and Bilal Piot and Koray Kavukcuoglu and R{\'e}mi Munos and Michal Valko},
  journal={ArXiv},
  year={2020},
  volume={abs/2006.07733},
  url={https://api.semanticscholar.org/CorpusID:219687798}
}

@article{Fang_2023,
doi = {10.3847/1538-3881/aca1a6},
url = {https://dx.doi.org/10.3847/1538-3881/aca1a6},
year = {2023},
month = {jan},
publisher = {The American Astronomical Society},
volume = {165},
number = {2},
pages = {35},
author = {GuanWen Fang and Shuo Ba and Yizhou Gu and Zesen Lin and Yuejie Hou and Chenxin Qin and Chichun Zhou and Jun Xu and Yao Dai and Jie Song and Xu Kong},
title = {Automatic Classification of Galaxy Morphology: A Rotationally-invariant Supervised Machine-learning Method Based on the Unsupervised Machine-learning Data Set},
journal = {The Astronomical Journal}
}

@article{Brammer_2008,
doi = {10.1086/591786},
url = {https://dx.doi.org/10.1086/591786},
year = {2008},
month = {oct},
publisher = {},
volume = {686},
number = {2},
pages = {1503},
author = {Gabriel B. Brammer and Pieter G. van Dokkum and Paolo Coppi},
title = {EAZY: A Fast, Public Photometric Redshift Code},
journal = {The Astrophysical Journal}
}

@article{Ilbert_2009,
doi = {10.1088/0004-637X/690/2/1236},
url = {https://dx.doi.org/10.1088/0004-637X/690/2/1236},
year = {2008},
month = {dec},
publisher = {The American Astronomical Society},
volume = {690},
number = {2},
pages = {1236},
author = {O. Ilbert and P. Capak and M. Salvato and H. Aussel and H. J. McCracken and D. B. Sanders and N. Scoville and J. Kartaltepe and S. Arnouts and E. Le Floc'h and B. Mobasher and Y. Taniguchi and F. Lamareille and A. Leauthaud and S. Sasaki and D. Thompson and M. Zamojski and G. Zamorani and S. Bardelli and M. Bolzonella and A. Bongiorno and M. Brusa and K. I. Caputi and C. M. Carollo and T. Contini and R. Cook and G. Coppa and O. Cucciati and S. de la Torre and L. de Ravel and P. Franzetti and B. Garilli and G. Hasinger and A. Iovino and P. Kampczyk and J.-P. Kneib and C. Knobel and K. Kovac and J. F. Le Borgne and V. Le Brun and O. Le Fèvre and S. Lilly and D. Looper and C. Maier and V. Mainieri and Y. Mellier and M. Mignoli and T. Murayama and R. Pellò and Y. Peng and E. Pérez-Montero and A. Renzini and E. Ricciardelli and D. Schiminovich and M. Scodeggio and Y. Shioya and J. Silverman and J. Surace and M. Tanaka and L. Tasca and L. Tresse and D. Vergani and E. Zucca},
title = {COSMOS PHOTOMETRIC REDSHIFTS WITH 30-BANDS FOR 2-deg2},
journal = {The Astrophysical Journal}
}

@article{Calzetti_2000,
doi = {10.1086/308692},
url = {https://dx.doi.org/10.1086/308692},
year = {2000},
month = {apr},
publisher = {},
volume = {533},
number = {2},
pages = {682},
author = {Daniela Calzetti and Lee Armus and Ralph C. Bohlin and Anne L. Kinney and Jan Koornneef and Thaisa Storchi-Bergmann},
title = {The Dust Content and Opacity of Actively
Star-forming Galaxies*},
journal = {The Astrophysical Journal}
}

@article{Peng_2002,
doi = {10.1086/340952},
url = {https://dx.doi.org/10.1086/340952},
year = {2002},
month = {jul},
publisher = {},
volume = {124},
number = {1},
pages = {266},
author = {Chien Y. Peng and Luis C. Ho and Chris D. Impey and Hans-Walter Rix},
title = {Detailed Structural Decomposition of Galaxy
Images*},
journal = {The Astronomical Journal}
}

@article{Lotz_2004,
doi = {10.1086/421849},
url = {https://dx.doi.org/10.1086/421849},
year = {2004},
month = {jul},
publisher = {},
volume = {128},
number = {1},
pages = {163},
author = {Jennifer M. Lotz and Joel Primack and Piero Madau},
title = {A New Nonparametric Approach to Galaxy Morphological Classification},
journal = {The Astronomical Journal}
}

@article{Conselice_2000,
doi = {10.1086/308300},
url = {https://dx.doi.org/10.1086/308300},
year = {2000},
month = {feb},
publisher = {},
volume = {529},
number = {2},
pages = {886},
author = {Christopher J. Conselice and Matthew A. Bershady and Anna Jangren},
title = {The Asymmetry of Galaxies: Physical Morphology for Nearby and High-Redshift Galaxies},
journal = {The Astrophysical Journal}
}

@article{Conselice_2003,
doi = {10.1086/375001},
url = {https://dx.doi.org/10.1086/375001},
year = {2003},
month = {jul},
publisher = {},
volume = {147},
number = {1},
pages = {1},
author = {Christopher J. Conselice},
title = {The Relationship between Stellar Light Distributions of Galaxies and
Their Formation Histories},
journal = {The Astrophysical Journal Supplement Series}
}

@article{10.1093/mnras/stt1016,
    author = {Freeman, P. E. and Izbicki, R. and Lee, A. B. and Newman, J. A. and Conselice, C. J. and Koekemoer, A. M. and Lotz, J. M. and Mozena, M.},
    title = "{New image statistics for detecting disturbed galaxy morphologies at high redshift}",
    journal = {Monthly Notices of the Royal Astronomical Society},
    volume = {434},
    number = {1},
    pages = {282-295},
    year = {2013},
    month = {06},
    issn = {0035-8711},
    doi = {10.1093/mnras/stt1016},
    url = {https://doi.org/10.1093/mnras/stt1016},
    eprint = {https://academic.oup.com/mnras/article-pdf/434/1/282/18499355/stt1016.pdf},
}

@article{Lotz_2006,
doi = {10.1086/497950},
url = {https://dx.doi.org/10.1086/497950},
year = {2006},
month = {jan},
publisher = {},
volume = {636},
number = {2},
pages = {592},
author = {Jennifer M. Lotz and Piero Madau and Mauro Giavalisco and Joel Primack and Henry C. Ferguson},
title = {The Rest-Frame Far-Ultraviolet Morphologies of Star-Forming Galaxies at z ~ 1.5 and 4},
journal = {The Astrophysical Journal}
}

@InProceedings{10.1007/978-3-642-21735-7_7,
author="Masci, Jonathan
and Meier, Ueli
and Cire{\c{s}}an, Dan
and Schmidhuber, J{\"u}rgen",
editor="Honkela, Timo
and Duch, W{\l}odzis{\l}aw
and Girolami, Mark
and Kaski, Samuel",
title="Stacked Convolutional Auto-Encoders for Hierarchical Feature Extraction",
booktitle="Artificial Neural Networks and Machine Learning -- ICANN 2011",
year="2011",
publisher="Springer Berlin Heidelberg",
address="Berlin, Heidelberg",
pages="52--59",
isbn="978-3-642-21735-7"
}

@article{Dai_2023,
doi = {10.3847/1538-4365/ace69e},
url = {https://dx.doi.org/10.3847/1538-4365/ace69e},
year = {2023},
month = {sep},
publisher = {The American Astronomical Society},
volume = {268},
number = {1},
pages = {34},
author = {Yao Dai and Jun Xu and Jie Song and Guanwen Fang and Chichun Zhou and Shuo Ba and Yizhou Gu and Zesen Lin and Xu Kong},
title = {The Classification of Galaxy Morphology in the H Band of the COSMOS-DASH Field: A Combination-based Machine-learning Clustering Model},
journal = {The Astrophysical Journal Supplement Series}
}

@misc{liu2023simple,
      title={Simple but Effective Unsupervised Classification for Specified Domain Images: A Case Study on Fungi Images}, 
      author={Zhaocong Liu and Fa Zhang and Lin Cheng and Huanxi Deng and Xiaoyan Yang and Zhenyu Zhang and Chichun Zhou},
      year={2023},
      eprint={2311.08995},
      archivePrefix={arXiv},
      primaryClass={cs.CV}
}

@article{MACKIEWICZ1993303,
title = {Principal components analysis (PCA)},
journal = {Computers {\&} Geosciences},
volume = {19},
number = {3},
pages = {303-342},
year = {1993},
issn = {0098-3004},
doi = {https://doi.org/10.1016/0098-3004(93)90090-R},
url = {https://www.sciencedirect.com/science/article/pii/009830049390090R},
author = {Andrzej Maćkiewicz and Waldemar Ratajczak}
}

@inproceedings{10.1145/233269.233324,
author = {Zhang, Tian and Ramakrishnan, Raghu and Livny, Miron},
title = {BIRCH: an efficient data clustering method for very large databases},
year = {1996},
isbn = {0897917944},
publisher = {Association for Computing Machinery},
address = {New York, NY, USA},
url = {https://doi.org/10.1145/233269.233324},
doi = {10.1145/233269.233324},
booktitle = {Proceedings of the 1996 ACM SIGMOD International Conference on Management of Data},
pages = {103–114},
numpages = {12},
location = {Montreal, Quebec, Canada},
series = {SIGMOD '96}
}

@Inbook{Peng2020,
author="Peng, Kai
and Zhang, Yiwen
and Wang, Xiaofei
and Xu, Xiaolong
and Li, Xiuhua
and Leung, Victor C. M.",
title="Computation Offloading in Mobile Edge Computing",
bookTitle="Encyclopedia of Wireless Networks",
year="2020",
publisher="Springer International Publishing",
address="Cham",
pages="216--220",
isbn="978-3-319-78262-1",
doi="10.1007/978-3-319-78262-1_331",
url="https://doi.org/10.1007/978-3-319-78262-1_331"
}

@article{10.1093/comjnl/26.4.354,
    author = {Murtagh, F.},
    title = "{A Survey of Recent Advances in Hierarchical Clustering Algorithms}",
    journal = {The Computer Journal},
    volume = {26},
    number = {4},
    pages = {354-359},
    year = {1983},
    month = {11},
    issn = {0010-4620},
    doi = {10.1093/comjnl/26.4.354},
    url = {https://doi.org/10.1093/comjnl/26.4.354},
    eprint = {https://academic.oup.com/comjnl/article-pdf/26/4/354/1072603/26-4-354.pdf},
}

@article{Murtagh_2014,
   title={Ward’s Hierarchical Agglomerative Clustering Method: Which Algorithms Implement Ward’s Criterion?},
   volume={31},
   ISSN={1432-1343},
   url={http://dx.doi.org/10.1007/s00357-014-9161-z},
   DOI={10.1007/s00357-014-9161-z},
   number={3},
   journal={Journal of Classification},
   publisher={Springer Science and Business Media LLC},
   author={Murtagh, Fionn and Legendre, Pierre},
   year={2014},
   month=oct, pages={274–295} }

@article{8ddb7f85-9a8c-3829-b04e-0476a67eb0fd,
 ISSN = {00359254, 14679876},
 URL = {http://www.jstor.org/stable/2346830},
 author = {J. A. Hartigan and M. A. Wong},
 journal = {Journal of the Royal Statistical Society. Series C (Applied Statistics)},
 number = {1},
 pages = {100--108},
 publisher = {[Wiley, Royal Statistical Society]},
 title = {Algorithm AS 136: A K-Means Clustering Algorithm},
 urldate = {2024-03-13},
 volume = {28},
 year = {1979}
}

@misc{huang2025unsupervisedwasteclassificationdualencoder,
      title={Unsupervised Waste Classification By Dual-Encoder Contrastive Learning and Multi-Clustering Voting (DECMCV)}, 
      author={Kui Huang and Mengke Song and Shuo Ba and Ling An and Huajie Liang and Huanxi Deng and Yang Liu and Zhenyu Zhang and Chichun Zhou},
      year={2025},
      eprint={2503.02241},
      archivePrefix={arXiv},
      primaryClass={cs.CV},
      url={https://arxiv.org/abs/2503.02241}, 
}

@article{Yao_2023,
doi = {10.3847/1538-4357/ace7b5},
url = {https://dx.doi.org/10.3847/1538-4357/ace7b5},
year = {2023},
month = {aug},
publisher = {The American Astronomical Society},
volume = {954},
number = {2},
pages = {113},
author = {Yao Yao and Jie Song and Xu Kong and Guanwen Fang and Hong-Xin Zhang and Xinkai Chen},
title = {Evolution of Nonparametric Morphology of Galaxies in the JWST CEERS Field at z ≃ 0.8–3.0},
journal = {The Astrophysical Journal}
}

@article{10.1111/j.1365-2966.2007.12627.x,
    author = {Ball, N. M. and Loveday, J. and Brunner, R. J.},
    title = "{Galaxy colour, morphology and environment in the Sloan Digital Sky Survey}",
    journal = {Monthly Notices of the Royal Astronomical Society},
    volume = {383},
    number = {3},
    pages = {907-922},
    year = {2008},
    month = {01},
    issn = {0035-8711},
    doi = {10.1111/j.1365-2966.2007.12627.x},
    url = {https://doi.org/10.1111/j.1365-2966.2007.12627.x},
    eprint = {https://academic.oup.com/mnras/article-pdf/383/3/907/3470252/mnras0383-0907.pdf},
}

@article{Gu_2018,
doi = {10.3847/1538-4357/aaad0b},
url = {https://dx.doi.org/10.3847/1538-4357/aaad0b},
year = {2018},
month = {feb},
publisher = {The American Astronomical Society},
volume = {855},
number = {1},
pages = {10},
author = {Yizhou Gu and Guanwen Fang and Qirong Yuan and Zhenyi Cai and Tao Wang},
title = {The Morphological Evolution, AGN Fractions, Dust Content, Environments, and Downsizing of Massive Green Valley Galaxies at 0.5 &lt; z &lt; 2.5 in 3D-HST/CANDELS},
journal = {The Astrophysical Journal}
}

@article{Barden2012GALAPAGOSFP,
  title={GALAPAGOS: From Pixels to Parameters},
  author={Marco Barden and Boris Haussler and Chien Y. Peng and D. H. Mcintosh and Yicheng Guo},
  journal={Monthly Notices of the Royal Astronomical Society},
  year={2012},
  volume={422},
  pages={449-468},
  url={https://api.semanticscholar.org/CorpusID:119117461}
}

@article{Song_2024,
doi = {10.3847/1538-4365/ad434f},
url = {https://dx.doi.org/10.3847/1538-4365/ad434f},
year = {2024},
month = {jun},
publisher = {The American Astronomical Society},
volume = {272},
number = {2},
pages = {42},
author = {Jie Song and GuanWen Fang and Shuo Ba and Zesen Lin and Yizhou Gu and Chichun Zhou and Tao Wang and Cai-Na Hao and Guilin Liu and Hongxin Zhang and Yao Yao and Xu Kong},
title = {USmorph: An Updated Framework of Automatic Classification of Galaxy Morphologies and Its Application to Galaxies in the COSMOS Field},
journal = {The Astrophysical Journal Supplement Series}
}

@ARTICLE{2022A&A...664A..92H,
       author = {{H{\"a}u{\ss}ler}, Boris and {Vika}, Marina and {Bamford}, Steven P. and {Johnston}, Evelyn J. and {Brough}, Sarah and {Casura}, Sarah and {Holwerda}, Benne W. and {Kelvin}, Lee S. and {Popescu}, Cristina},
        title = "{GALAPAGOS-2/GALFITM/GAMA - Multi-wavelength measurement of galaxy structure: Separating the properties of spheroid and disk components in modern surveys}",
      journal = {\aap},
         year = 2022,
        month = aug,
       volume = {664},
          eid = {A92},
        pages = {A92},
          doi = {10.1051/0004-6361/202142935},
archivePrefix = {arXiv},
       eprint = {2204.05907},
 primaryClass = {astro-ph.GA},
       adsurl = {https://ui.adsabs.harvard.edu/abs/2022A&A...664A..92H}
}

@INPROCEEDINGS{7298594,
  author={Szegedy, Christian and Wei Liu and Yangqing Jia and Sermanet, Pierre and Reed, Scott and Anguelov, Dragomir and Erhan, Dumitru and Vanhoucke, Vincent and Rabinovich, Andrew},
  booktitle={2015 IEEE Conference on Computer Vision and Pattern Recognition (CVPR)}, 
  title={Going deeper with convolutions}, 
  year={2015},
  volume={},
  number={},
  pages={1-9},
  doi={10.1109/CVPR.2015.7298594}}
\bibliographystyle{aasjournal}

\movetabledown=20mm  
\begin{rotatetable*}
\begin{deluxetable*}{ccccccccccccccccccccccccccccc}
\centerwidetable
\tablecaption{Part of The Final Catalogue \label{tab5}}
\tablehead{
  \colhead{Seq.} & 
  \colhead{R.A.} & 
  \colhead{DEC.} & 
  \colhead{$M_{\ast}$} & 
  \colhead{z} & 
  \colhead{$r_e$} & 
  \colhead{n} & 
  \colhead{$\rm flag1$} & 
  \colhead{C} & 
  \colhead{A} & 
  \colhead{S} & 
  \colhead{G} & 
  \colhead{$M_{20}$} & 
  \colhead{M} & 
  \colhead{I} & 
  \colhead{D} & 
  \colhead{$G_{2}$} & 
  \colhead{$\Psi$} & 
  \colhead{$\rm flag2$} & 
  \colhead{Type} & 
  \colhead{Label} \\
  & (deg) & (deg) & ($\log M_{\odot}$) &  & (kpc) & & & & & & & & & & & & & & & \\ 
(1) & (2) & (3) & (4) & (5) & (6) & (7) & (8) & (9) & (10) & (11) & (12) & (13) & (14) & (15) & (16) & (17) & (18) & (19) & (20) & (21)\\}
\startdata  
1  & 149.74180 & 2.14655 & 9.48  & 0.92  & 4.00  & 0.54  & 0 & 2.48  & 0.06  & 0.02  & 0.41  & -1.29 & 0.15 & 0.38 & 0.26  & 2.00  & 3.45  & 0  & 7 & IRR\\
2 & 150.24595 & 1.81537 & 9.54 & 0.66 & 4.19 & 0.83 & 0 & 2.60 & 0.10 & 0.09 & 0.50 & -1.54 & 0.21 & 0.39 & 0.05 & 2.01 & 3.90 & 0 & 11 & IRR\\
3  & 149.76172 & 2.17835 & 9.91  & 0.85  & 4.16  & 0.64  & 0 & 2.32  & 0.07  & 0.03  & 0.42  & -1.32 & 0.32 & 0.46  & 0.11  & 2.01  & 2.96  & 0  & 7 & IRR\\
4  & 149.76514 & 2.16217 & 8.67  & 0.73  & 1.75  & 0.71  & 0 & 2.61  & 0.11  & 0.02  & 0.50  & -1.34 & 0.71 & 0.66  & 0.14  & 1.98  & 2.19  & 0  & 6 & ETD \\
5  & 149.76802 & 2.20867 & 9.00  & 0.78  & 4.69  & 0.90  & 0 & 3.06  & -0.07 & 0.00  & 0.50  & -1.42 & 0.26 & 0.32  & 0.12  & 2.01  & 8.53  & 0  & 1 & LTD \\
6  & 150.35354 & 2.04999 & 9.83  & 0.50  & 3.60  & 0.90  & 0 & 2.76  & 0.02  & 0.02  & 0.48  & -1.70 & 0.06 & 0.19  & 0.07  & 2.00  & 2.72  & 0  & 13 & LTD\\
7  & 149.88250 & 2.18163 & 9.92  & 0.91  & 3.99  & 1.08  & 0 & 3.10  & -0.10 & 0.02  & 0.46  & -1.53 & 0.17 & 0.40  & 0.13  & 2.01  & 6.15  & 0  & 1 & LTD\\
8  & 149.78566 & 2.27533 & 9.12  & 1.13  & 3.89  & 1.27  & 0 & 2.61  & -0.01 & 0.03  & 0.47  & -1.51 & 0.18 & 0.16  & 0.08  & 2.01  & 4.25  & 0  & 2 &LTD\\
9  & 150.20156 & 2.03222 & 9.43  & 2.34  & 1.78  & 3.24  & 0 & 3.50  & -0.16 & 0.00  & 0.54  & -1.84 & 0.22 & 0.11  & 0.18  & 2.01  & 3.64  & 0  & 4 & SPH \\
10 & 150.10518 & 1.90293 & 9.09  & 1.08  & 4.81  & 0.95  & 0 & 2.31  & -0.10 & 0.05  & 0.46  & -1.20 & 0.29 & 0.46  & 0.51  & 2.03  & 7.52  & 0  & 9& IRR \\
11 & 150.21963 & 1.77530 & 9.45  & 1.00  & 4.28  & 0.72  & 0 & 2.14  & 0.02  & 0.01  & 0.45  & -1.11 & 0.62 & 0.63  & 0.06  & 2.01  & 3.78  & 0  & 9 & IRR\\
12 & 150.20432 & 2.10398 & 9.74  & 3.27  & 2.06  & 1.62  & 0 & 2.91  & 0.10  & 0.01  & 0.50  & -1.48 & 0.03 & 0.02  & 0.21  & 2.00  & 2.26  & 0  & 3 & ETD\\
13 & 150.17316 & 1.90306 & 9.43  & 1.20  & 1.64  & 3.14  & 0 & 3.90  & -0.04 & 0.01  & 0.62  & -2.12 & 0.07 & 0.01  & 0.01  & 1.92  & 2.00  & 0  & 0& SPH \\
14 & 150.21701 & 1.81433 & 8.94  & 0.56  & 1.74  & 3.86  & 0 & 3.71  & 0.03  & -0.01 & 0.59  & -2.03 & 0.11 & 0.13  & 0.06  & 1.81  & 2.03  & 0  & 0 & SPH\\
15 & 150.25411 & 1.79772 & 9.98 & 1.86 & 1.49 & 3.16 & 0 & 3.39 & 0.02 & 0.03 & 0.60 & -1.80 & 0.04 & 0.02 & 0.09 & 1.95 & 1.64 & 0 & 14 & SPH\\
16 & 149.84097 & 2.21220 & 9.47 & 2.98 & 2.02 & 1.49 & 0 & 2.85 & -0.01 & 0.01 & 0.49 & -1.75 & 0.03 & 0.01 & 0.04 & 2.00 & 1.82 & 0 & 12 & ETD\\
17 & 149.76096 & 2.02256 & 9.11  & 0.87  & 1.89  & 1.71  & 0 & 3.06  & 0.03  & 0.00  & 0.50  & -1.72 & 0.06 & 0.02  & 0.09  & 1.95  & 2.54  & 0  & 6 & ETD\\
18 & 150.24429 & 1.99855 & 8.45  & 0.65  & 1.94  & 0.51  & 0 & 2.22  & -0.04 & 0.04  & 0.40  & -1.26 & 0.17 & 0.30  & 0.11  & 2.01  & 2.81  & 0  & 5& UNC \\
19 & 150.14808 & 2.03992 & 9.20 & 0.83 & 2.53 & 0.56 & 0 & 2.18 & 0.12 & 0.03 & 0.44 & -1.23 & 0.21 & 0.19 & 0.09 & 2.01 & 1.90 & 0 & 10 & UNC \\
20 & 150.30796 & 1.99565 & 9.44  & 2.85  & 1.83  & 1.39  & 0 & 3.22  & -0.04 & 0.02  & 0.49  & -1.24 & 0.12 & 0.24  & 0.31  & 2.01  & 4.29  & 0  & 8& UNC \\
\enddata
\tablecomments{The column (1) is the sequential number identifier. The columns (2) and (3) correspond to the R.A and DEC. in degree. The columns (4) and (5) shows stellar masses and redshifts in the I band \citep{Weaver_2022}, respectively. Columns (6) and (7) list parametric morphological measurements, including the effective radius ($r_e$) and S\'{e}rsic index ($n$). The flag1 in column (8) represents the goodness of  single S\'{e}rsic fitting, and the flag1=0 means a good fitting. Columns (9)-(18) show the non-parametric morphological measurements, including C, A, S, Gini, $M_{\rm 20}$, M, I, D, $G_2$, and $\Psi$ coefficient, respectively.  The flag2 in column (19) represents the goodness of the nonparametric morphological measurements, where flag2=0 means a good measurement; Columns (20) and (21) show the information about the morphological classification: 0, 4, 14 for SPH; 3, 6, 12 for ETD; 1, 2, 13 for LTD; 7, 9, 11 for IRR; 5, 8, 10 for UNC. Labels 0-9 were obtained from UML, while labels 10-14 were obtained from SML.}  
\end{deluxetable*}
\end{rotatetable*}

\end{document}